\documentclass[aps,pra,showpacs,twocolumn,eqsecnum]{revtex4}
\usepackage[dvips,final]{graphicx}
\usepackage{bm, graphics, amsmath}

\begin{document}

\title{Optimal state discrimination with a fixed rate of inconclusive results: \\ Analytical solutions and relation to state discrimination with a fixed error rate  }

\author{Ulrike Herzog}
\affiliation {Nano-Optics, Institut f\"ur Physik,
Humboldt-Universit\"at Berlin, Newtonstrasse 15, D-12489 Berlin,
Germany}

\begin{abstract}
We study an optimum  measurement for quantum state discrimination, which  maximizes the probability of correct results when the probability of inconclusive results is fixed at a given value. 
 The measurement  describes  minimum-error discrimination if this value is zero, while under certain conditions it corresponds to optimized maximum-confidence discrimination, or to optimum unambiguous discrimination, respectively,  when the fixed value reaches a definite minimum. 
Using operator conditions that determine the optimum measurement, we derive  analytical solutions for the discrimination of two mixed qubit states, including the case of two pure states occurring with arbitrary prior probabilities, and for  the discrimination of  $N$ symmetric states,
 both pure and mixed. We also consider a case where the given density operators  resolve the identity operator, and we specify the optimality conditions for  partially symmetric states.  Moreover, we show  that from  the complete solution 
for arbitrary values of the fixed rate of inconclusive results one can always  obtain  the optimum measurement in another strategy where the  error rate is fixed, and vice versa.  
\end{abstract}


\maketitle

\section{Introduction }
In quantum state discrimination we want to determine the actual state of a quantum system that is known to be in a certain state belonging to a given set of states. This is an essential problem for many tasks in quantum communication and quantum cryptography.
Since nonorthogonal quantum states cannot be distinguished
perfectly, various optimized discrimination strategies have been
developed.
The  best known of these are minimum-error discrimination  
  \cite{helstrom, holevo} 
and optimum unambiguous discrimination, where the latter strategy was originally derived for pure states  \cite{ivan,dieks,peres,jaeger} and later also considered for mixed states \cite{rudolph,raynal,eldar,HB,BFH,herzog,raynal2,kleinmann}.
In unambiguous discrimination  errors do not occur, that is, the total error rate $P_e$  is required to vanish, $P_e =0$.
This can be achieved at the expense of admitting inconclusive results, where the measurement fails to give a definite answer. Optimum unambiguous discrimination minimizes this failure probability,  or failure rate, $Q$, yielding the minimum value $Q=Q_{min}$.  

Unambiguous state discrimination is not possible for pure states that are linearly dependent \cite{chefles1}, or for mixed states described by density operators with identical  supports  \cite{rudolph,raynal}, where the support is the Hilbert space spanned by the  eigenvectors with non-zero eigenvalues. When unambiguous discrimination is impossible, related measurements can be applied, which discriminate the states with  maximum possible confidence for each
 conclusive  outcome  \cite{croke,mosley}.
Optimized maximum-confidence discrimination is achieved by the particular one of these  measurements that minimizes the failure probability \cite{herzog1,herzog-benson,steudle,jimenez,herzog2}. This measurement  corresponds to optimum unambiguous discrimination  when for each of the different conclusive results  the  maximum confidence is equal to unity.

While the confidence is defined separately for each conclusive outcome, in minimum-error discrimination   the overall error rate $P_e$, averaged over all outcomes, is minimized.
In this strategy inconclusive results are not allowed, that is, $Q=0$.
Chefles and Barnett \cite{chefles-barnett} introduced a more general strategy that minimizes $P_e$ when a certain fixed rate  $Q$ of inconclusive results  is admitted, thus reducing the minimum achievable value of $P_e$. For discriminating two equiprobable nonorthogonal pure states these authors  obtained the optimum solution, which interpolates  between minimum-error discrimination and optimum  unambiguous discrimination when the fixed rate $Q$ grows from $Q=0$ to $Q=Q_{min}$.  Their research triggered further investigations \cite{zhang-li,fiurasek,eldar1,touzel,hayashi-err,sugimoto}. 
In particular,   optimum state discrimination with a fixed  
 rate  $Q$ of inconclusive results was extended to mixed states
\cite{fiurasek,eldar1}, and the relation to the strategy of maximum-confidence discrimination  was briefly discussed in our previous papers \cite{herzog1,herzog2}. 
 Clearly, in a measurement where $Q$ is fixed, minimizing  $P_e$ corresponds to maximizing the overall rate of correct results, $P_c$.
  Fiur\'a\v{s}ek and Je\v{z}ek  \cite{fiurasek} derived general operator conditions that have to be fulfilled in the optimum measurement with an arbitrary fixed  value $Q$, holding for the discrimination of an arbitrary number $N$ of mixed states. For $N=2$ they solved the optimization problem 
 in the special case of two mixed qubit that have the same purity and occur with equal prior probabilities \cite{fiurasek}, which includes the solution for two equiprobable pure states \cite{chefles-barnett}. Other solutions were not obtained.   

Recently optimum state discrimination has been also investigated for a measurement where a certain fixed error rate $P_e$ is admitted \cite{touzel,hayashi-err,sugimoto}.
With a fixed rate of $P_e$, maximizing  the  rate of correct results $P_c$ corresponds to minimizing the failure rate $Q$. The optimum measurement in this strategy was studied for the discrimination of two states, $N=2$, where the solution was derived for two pure states occurring with arbitrary prior probabilities  \cite{hayashi-err,sugimoto}. However, it was not recognized that there exists a relation between optimum state discrimination with a fixed value of $Q$, on the one hand, and with a fixed value of $P_e$, on the other hand. This relation implies that by completely solving the optimization problem for one of these two discrimination strategies one can also obtain the solution for the other strategy, as will be shown in the present paper.   

The purpose of  our paper is twofold. First, for optimum state discrimination with a fixed rate of $Q$ we derive analytical solutions going beyond the solutions obtained so far \cite{chefles-barnett,fiurasek},
and we also outline the relation to maximum-confidence discrimination, extending our 
previous discussions \cite{herzog1,herzog2}. 
Second,  we show that when the maximum  rate of correct results $P_c$ is known as a function of the fixed failure probability $Q$, then from this function one can also obtain the maximum  rate of correct results in dependence of a fixed  error rate $P_e$, and vice versa, due to a general relation between the solutions of the two optimized strategies that holds for the discrimination of an arbitrary number $N$ of mixed states.  

The paper is organized as follows: We begin  in Sec. II with an alternative derivation of the optimality conditions for state discrimination with fixed $Q$, and with establishing the relation to maximum-confidence discrimination.
In Sec. III  together with Appendix  A we apply these conditions to the discrimination  of  two mixed qubit states occurring with arbitrary prior probabilities, while Sec. IV together with Appendix B  is devoted to the discrimination of $N$ symmetric states.   In Sec. V we consider a case where the density operators  resolve the identity operator, and we also specify the optimality conditions for a  case of partially symmetric states. The general relation between the two optimization strategies where either $P_e$ or $Q$ has a fixed value is presented in Sec. VI, together with an example.  Sec. VII provides a summary of results and concludes the paper. 

\section{ Optimum measurement}

\subsection{Conditions for optimality}

We suppose that a quantum system is  prepared with the prior
probability $\eta_j$ in one  of $N$ given states described by the
density operators $\rho_j$ $(j=1,\ldots,N)$, where 
$\sum_{j=1}^N \eta_j =1$. 
The general task is to perform a discrimination measurement
in order to infer in which of the $N$ possible
states the system was prepared.
 The  measurement is  described by $N+1$ positive detection
 operators $\Pi_1,\ldots \Pi_N$ and $\Pi_0$,   
where ${\rm Tr\,\,\,}(\rho_k\Pi_j)$ is the conditional probability that a system
 is inferred to be in the state $\rho_j$  given it had been prepared
 in the state $\rho_k$, while ${\rm Tr\,}(\rho_k\Pi_0)$
is the conditional probability that in this case an
inconclusive result is obtained and the measurement fails to discriminate the states.
The detection operators fulfill the completeness relation
$\sum_{j=0}^N \Pi_j=I$, where $I$ is the identity operator in the  $d$-dimensional  Hilbert space
${\cal H}_d$ spanned by the eigenstates of the operators $\rho_1,\ldots, \rho_N$ that belong to nonzero eigenvalues.  Unless each of the detection operators is a projector, the measurement is a generalized one.  Once the detection operators are known, implementations of the  generalized measurement as a projective measurement in an enlarged Hilbert space can be obtained using standard methods \cite{neumark,preskill}. 
 
We are interested in the specific discrimination measurement where the  overall failure probability 
\begin{equation}
\label{Q}
Q=\sum_{j=1}^N\eta_j{\rm Tr\,} (\rho_j\Pi_0)= {\rm Tr\,}(\rho \Pi_0) \quad{\rm with}\;\;  \rho=  \sum_{j=1}^N \eta_j \rho_j 
\end{equation}
has a given fixed value $Q<1$, while the  overall  probability of getting  correct results,  $P_c$, 
is as large as possible. Here we defined  the total density operator $\rho$, which  has its support in the full Hilbert space  ${\cal H}_d$ and is thus an operator of rank $d$. 
Our optimization problem can be expressed as follows:
\begin{eqnarray}
\label {Primal1}&&{\rm maximize}\; P_c  =\! \sum_{j=1}^N \!\eta_j{\rm Tr\,}(\rho
_j\Pi_j)
=1-Q-P_e,
\qquad\\
\label{Qprimal1a}
&& \mbox{subject to}\;\;
Q={\rm Tr\,}(\rho \Pi_0) ={\rm const},\quad
 \sum_{j=0}^N\Pi_j = I.\qquad
\end{eqnarray}
 In Eq. (\ref{Primal1}) we introduced the overall error probability $P_e$. 
Since $Q$ is fixed, the maximum  of the absolute rate of correct results, $P_c$,  determines also the maximum of the relative rate of correct results at the same value of $Q$,  defined as $R_c= P_c/(1-Q)$ \cite{fiurasek}. We thus get  
\begin{equation}
\label{rel-rate} 
P_c^{max}\big|_Q =  R_c^{max} \big|_Q (1-Q).
\end{equation}
In the special case  where $R_c^{max}|_Q=1$ errors do not occur and the states are unambiguously discriminated. 

As will be shown in Sec. II B,  if  $Q$ exceeds  a certain minimum, denoted by $Q^{\prime}$, the maximum relative rate of correct results, $R_c^{max}|_Q$, stays constant with growing $Q$.
 The optimum measurement resulting from Eqs. (\ref{Primal1}) and (\ref{Qprimal1a})  with  $Q >  Q^{\prime}$  then  yields a maximum of $P_c$ that is smaller than  $P_c^{max}|_{Q^{\prime}}$ and is therefore without practical relevance. One could modify the optimization problem and ask for the maximum of $P_c$ under the constraint that $Q$ does not exceed a certain fixed margin, $Q_M$. The  maximum of $P_c$ under this modified constraint then follows from the solution of the original maximization  problem,  Eqs.   (\ref{Primal1}) and (\ref{Qprimal1a}), and is given by   $P_c^{max}|_{Q_M}$ if $Q_M \leq Q^{\prime}$ and  by $P_c^{max}|_{Q^{\prime}}$ if  $Q_M \geq Q^{\prime}$. 

In order to derive analytical solutions of the optimization problem posed by Eqs.  (\ref{Primal1}) and (\ref{Qprimal1a}) we use the operator conditions \cite{fiurasek,eldar1} that have to be fulfilled in the optimum measurement. 
Let us begin by re-deriving these optimality conditions. For this purpose we introduce a Hermitian operator $Z$ and a scalar real amplifier $a$. Due to the two constraints in Eq. (\ref{Qprimal1a})   the equation 
\begin{equation}
\label {Q1} {\rm Tr\,} Z -a Q- P_c= {\rm Tr\, }[(Z-a\rho)\Pi_0] + \sum_{j=1}^N
{\rm Tr\, }[(Z-\eta_j\rho_j)\Pi_j]
\end{equation}
is identically fulfilled for any operator $Z$ and any multiplier  $a$. 
Since the detection operators are positive it follows that the positivity conditions
\begin{equation}
\label {Q2}
Z- a \rho \geq 0, \quad Z-\eta_j\rho_j\geq 0
\end{equation}
$(j=1,\ldots, N)$ imply that   
${\rm Tr\,} Z -  a Q - P_c \geq0.$ 
Hence when the positivity conditions 
in Eq. (\ref{Q2}) are fulfilled, the minimum of ${\rm Tr\,} Z-a Q $  establishes an upper bound for $P_c$. 
In the optimum measurement,  where $P_c$ is equal to this bound, both sides of Eq. (\ref{Q1}) vanish
 for the  optimum multiplier $a$ and the optimum  operators  $Z$,   $\Pi_0$ and $\Pi_j$ $(j=1,\ldots,N)$.  
Due to Eq. (\ref{Q2}) this requires that each single term on   
 the right-hand side of Eq. (\ref{Q1}) vanishes separately, that is  
\begin{equation}
\label {Q4}
 (Z-a \rho)\Pi_0 =0, \quad (Z-\eta_j\rho_j)\Pi_j =0
\end{equation}
 $(j=1,\ldots,N)$.  Equations (\ref {Q2})  and (\ref {Q4}) together therefore establish sufficient optimality conditions, 
first derived by  Fiur\'a\v{s}ek and Je\v{z}ek  \cite{fiurasek} with the help of Lagrangian multipliers.  Using methods of semidefinite programming, the optimality conditions  have been shown to be not only sufficient,  but also necessary \cite{eldar1}. 
For solving the optimization problem 
it is sometimes advantageous to introduce transformed operators  \cite{fiurasek,croke,herzog1}. With 
\begin{eqnarray}
\label {s6a}
 \bar{\Pi}_0 &=& \rho^{1/2}\Pi_0  \rho^{1/2}, \quad  \Gamma=  \rho^{-1/2} Z \rho^{-1/2},\quad\\
\label {s6}
\bar{\Pi}_j &=& \rho^{1/2}\Pi_j  \rho^{1/2},\quad \tilde{\rho}_j=   \rho^{-1/2} \eta_j\rho_j   \rho^{-1/2}\quad 
 \end{eqnarray}
for $j=1,\ldots,N$,  Eqs. (\ref{Q2}) and  (\ref{Q4})  yield the  optimality conditions  
\begin{eqnarray}
\label {s9}
\! \Gamma- \tilde{\rho}_j \!& \geq& \! 0, \quad  (\Gamma  -\tilde{\rho}_j )\bar{\Pi}_j =0 \quad(j=1,\ldots,N),\qquad \\
\label {s8}
\Gamma \!- a I \!& \geq& \! 0, \quad (\Gamma- a I)\bar{\Pi}_0 =0, 
\end{eqnarray}
where   $\bar{\Pi}_0 +\sum_{j=1}^N\bar{\Pi}_j=\rho$, due to  the completeness relation of the detection operators. 
This representation of the optimality conditions results by multiplying Eqs.  (\ref{Q2}) and (\ref{Q4}) from the left and right by the Hermitian operator $\rho^{-1/2}$, taking into account that for any operator $A$ in ${\cal H}_d$ the relation $A  \rho^{-1}=0$ can only hold when $A=0$, since  the support  of $\rho$ is  the full  Hilbert space ${\cal H}_d$. 
 
Provided that an operator $Z$, a scalar multiplier $a$ and positive detection operators $ \Pi_1,\ldots \Pi_N$  with $\Pi_0 =I- \sum_{j=1}^N \geq 0$ satisfy Eqs. (\ref{Q2}) and  (\ref{Q4}), or Eqs. (\ref{s6a}) -- (\ref{s8}), respectively,  then the  detection operators determine the  optimum measurement, which  maximizes $P_c$ with the 
 fixed value $Q  =  {\rm Tr\,}(\rho\Pi_0) = {\rm Tr\,}  \bar{\Pi}_0$   and yields   
\begin{equation}
\label {Qdual5}
P_c^{max}\big|_Q \!=\!\sum_{j=1}^N \!\eta_j{\rm Tr\,}(\rho
_j\Pi_j)= \sum_{j=1}^N {\rm Tr\,\,}(\tilde{\rho} _j\bar{\Pi}_j). 
\end{equation}
By taking the trace in  both equalities in Eq. (\ref{Q4}) and summing over all states in the second equality we arrive at the expressions  \cite{fiurasek}
\begin{equation}
\label {Qdual5a}
{\rm Tr\,}(Z\Pi_0)=aQ,\quad P_c^{max}\big|_Q= {\rm Tr\,}[ Z(I-\Pi_0)] ={\rm Tr\,} Z-a Q. 
\end{equation}
If $Q=0$, that is, if inconclusive results are not allowed, the optimum measurement corresponds to minimum-error discrimination, described by the well-known optimality conditions arising from Eqs. (\ref{Q2}) and  (\ref{Q4}) with    $\Pi_0=0$ and $a=0$ \cite{helstrom,holevo}.

\subsection{Limiting case of sufficiently large $Q$ and relation to maximum-confidence discrimination} 

 When the fixed failure probability  $Q$ is getting larger and larger, the operator   $\bar{\Pi}_0$, defined in Eq. (\ref{s6a}), will turn into an operator of rank $d$ as soon as 
$Q$ reaches a certain minimum value $Q^{\prime}$ \cite{fiurasek},  due to the fact that for $Q=1$ we must have ${\Pi}_0=I$ and thus  $\bar{\Pi}_0=\rho$ with ${\rm rank}(\rho)=d$.  
 This means that for $Q\geq Q^{\prime}$ the operator   $\bar{\Pi}_0$ has its support in the full Hilbert space ${\cal H}_d$ and the equality in  Eq. (\ref{s8}) thus can  only hold when  $\Gamma - a I = 0$. For $Q\geq Q^{\prime}$ 
 the optimality conditions therefore reduce to   
\begin{equation}
\label {limit0}
aI-\tilde{\rho}_j \geq 0, \quad (aI-\tilde{\rho}_j )\bar{\Pi}_j =0\quad (j=1,\ldots,N)
 \end{equation}
\cite{fiurasek}. If $Q < 1$ at least one of the operators  $\bar{\Pi}_j$   has to be different from  zero. 
The equality condition in Eq. (\ref{limit0}) implies  that  for any $j$ where  $\bar{\Pi}_j\neq 0$  the eigenstates of   $aI \! -\!  \tilde{\rho}_j$ belonging to nonzero eigenvalues cannot span the full Hilbert space ${\cal H}_d$. Therefore when $\bar{\Pi}_j\neq 0$  at least one of the eigenvalues of $aI \! -\!  \tilde{\rho}_j$ is equal to zero in the optimum measurement with $Q\geq Q^{\prime}$.  Together with the positivity condition in  Eq. (\ref{limit0}) this requires that 
\begin{equation}
\label {lim1}
a=\mathrm{max} \{C_1,\ldots C_N\},\quad {\rm where}\;\;C_j ={\rm max}\{{\rm eig }(\tilde{\rho}_j) \},
 \end{equation}
that is, where   $C_j$ is the largest eigenvalue  of  $\tilde{\rho}_j$. For any state $j$ with $C_j < a$  the operator $aI-\tilde{\rho}_j$ has its support in the full Hilbert space and the equality in Eq. (\ref{limit0}) can only hold when   $\bar{\Pi}_j =0$.  For those states $j$ where  $\bar{\Pi}_j \neq 0$, the support of the operator $\bar{\Pi}_j $ is   the eigenspace of  $\tilde{\rho}_j$  belonging to its  largest eigenvalue,  since this guarantees that  $\bar{\Pi}_j $ is orthogonal to  $aI \! -\!  \tilde{\rho}_j$ \cite{fiurasek}. 
By taking the trace in  the equality in Eq. (\ref{limit0}),  summing over all states and inserting the value of $a$ we arrive at        
\begin{equation}
\label {lim2}
P_c^{max}\big|_Q
= \underset{\overset{j}{}}{\mathrm{max}} \{C_j\}(1-Q) 
\quad {\rm if} \;\;Q\geq Q^{\prime}.
 \end{equation}
In order to  determine  $Q^{\prime}$, we  have to minimize $Q=1-\sum_j{\rm Tr\,} \bar\Pi_j$ on  the conditions that the operators  $\bar\Pi_j$  have the required supports, as described after Eq. (\ref{lim1}), and that   
$\bar\Pi_0=\rho - \sum_{j=1}^N \bar\Pi_j \geq 0$.

The  eigenvalues  $C_j$ introduced in Eq. (\ref{lim1}) have a definite meaning in state discrimination. They  determine the maximum confidence  \cite{croke} that can be achieved  for the individual measurement outcome $j$  or, equivalently,    the maximum achievable ratio between all instances where the outcome $j$ is correct and  all instances where the outcome $j$ occurs. In fact, with Eq. (\ref {lim1}) it follows  that        
\begin{equation}
 \underset{\overset{ \Pi_j}{}}{\mathrm{max}} \left\{\frac{\eta_j{\rm Tr\,} (\rho_j\Pi_j)}{{\rm Tr\,} (\rho\Pi_j)}\right\} =
 \underset{\overset{\bar{\Pi}_j}{}}{\mathrm{max} }\left\{\frac{{\rm Tr\,} (\tilde{\rho}_j\bar{\Pi}_j)}{{\rm Tr\,} \bar{\Pi}_j}\right\} =
\label {conf} C_j, 
\end{equation}
where for each individual $j$ the maximization is performed with respect to all choices for the detection operator $\bar{\Pi}_j$, or  ${\Pi}_j$,   respectively. Clearly, in a measurement with  $C_j=1$ the state $j$ is unambiguously discriminated.
From Eq. (\ref{conf}) it becomes obvious that the maximum  confidence for the outcome $j$, equal to $C_j$,  is obtained when $\bar{\Pi}_j$ has its support in   the eigenspace of $\tilde{\rho}_j$  belonging to the eigenvalue $C_j$ \cite{croke,herzog1,herzog2}. When this condition holds for each of the $N$ states,  that is, when for each outcome $j$  the confidence is maximal,  the measurement is called a maximum-confidence measurement \cite{croke}. 

Optimized maximum-confidence discrimination is achieved by the specific maximum-confidence measurement where  
the probability of inconclusive results takes its smallest  possible value,  $Q_{min}^{MC}$ \cite{herzog1,herzog2}.
A comparison with the constant  $Q^{\prime}$ characterized after Eq.  
(\ref{lim2}) reveals  that  
\begin{equation}
\label {conf1}
  Q^{\prime}=Q_{min}^{MC}  \qquad{\rm if}\;\;C_1=\dots = C_N\equiv C.
\end{equation}
Hence in all cases where the maximum confidence is the same for each conclusive outcome,  the measurement  maximizing $P_c$ when $Q$ is fixed at the value $Q^{\prime}$ is equal to the measurement for optimized  maximum-confidence discrimination. When  $C=1$, the latter measurement  corresponds to optimum unambiguous discrimination. 

\section{Two mixed qubit states}

\subsection{Method for applying the optimality conditions}
 
In this paper we want to determine the optimum measurement for an arbitrary value of the fixed failure probability  $Q$, restricting ourselves to cases  that allow a simple analytical solution.  
Let us start with the  discrimination of two mixed qubit states in a joint two-dimensional Hilbert space, where we use the optimality conditions in the form of  Eqs. (\ref{s9}) and  (\ref{s8}). For   $N=2$ the  transformed density operators  $\tilde{\rho}_1$ and $\tilde{\rho}_2$  defined in Eq. (\ref{s6})  have identical systems of eigenstates, due to the relation $\tilde{\rho}_1+ \tilde{\rho}_2  = I$. Their spectral representations therefore can be written as   
\begin{eqnarray}
\label {s1a}     \tilde{\rho}_1 & =& C_1
|\nu_1 \rangle \langle \nu_1|+ (1-C_2) |\nu_2 \rangle \langle \nu_2|,\\
\label {s1b}
\tilde{\rho}_2& =&  (1-C_1) |\nu_1 \rangle \langle \nu_1|+ C_2 |\nu_2 \rangle \langle \nu_2|
 \end{eqnarray}
\cite{herzog1}, where $C_1> 1/2$ and  $C_2 > 1/2$, since the case $C_1=C_2=1/2$ would imply that $\rho_1= \rho_2$.    The constants $C_1$ and $C_2$ have the meaning of the maximum achievable confidence for the two respective outcomes, cf. Eqs. (\ref{lim1}) and (\ref{conf}). 
It is convenient to use the orthonormal eigenstates $|\nu_1\rangle$ and $|\nu_2\rangle$ as the basis states for solving the optimization problem.  For this purpose we define the   matrix elements
\begin{eqnarray}
\label {s2} 
 \langle \nu_j|\rho |\nu_j\rangle & =& \rho_{jj},\quad  \langle \nu_1|\rho |\nu_2\rangle  = \rho_{12}= | \rho_{12}|^{i\phi},
 \\
\label {s2a} 
\langle \nu_j|\Gamma|\nu_j\rangle&=&\Gamma_{jj},\quad \!\langle \nu_1|\Gamma|\nu_2\rangle= \Gamma_{12}=| \Gamma_{12}|e^{i\delta}\;\;
 \end{eqnarray}
 with $j=1,2$, where $\rho_{11}+\rho_{22}=1$.   In the special case when the two given states  are pure,   $\rho_j=|\psi_j\rangle\langle\psi_j|$,   we get  from Eq. (\ref{s6})  $ \tilde{\rho}_j= |\nu_j\rangle\langle \nu_j|$ with $\rho^{1/2}|\nu_j\rangle = \sqrt{\eta_j}|\psi_j\rangle$, which yields    
\begin{equation}
\label {s11}
 \rho_{12}=\sqrt{\eta_1\eta_2}\;\langle\psi_1|\psi_2\rangle, \quad \rho_{jj}=\eta_j, \quad C_j=1.
 \end{equation}
The relation $C_1=C_2=1$ reflects the fact that each of the two pure states can be unambiguously discriminated. 
Using Eq. (\ref{s6}) we obtain  $ {\rm Tr\,}(\rho \tilde{\rho}_j)= \eta_j $,  leading to  
\begin{equation}
\label {eta1}
  C_1\rho_{11}\!+\!(1\!-\!C_2)\,\rho_{22}\!=\! \eta_1,\quad   (1\!-\!C_1)\,\rho_{11}\!+\!C_2\rho_{22} \!=\!  \eta_2.
\end{equation}

Now we are prepared to apply the optimality conditions. 
We start with the  assumption  that for  the given value of $Q$ these  conditions are satisfied by a solution where 
$ \Gamma - a I > 0$ and where 
 all three detection operators are different from zero. 
 Equation (\ref{s8}) then requires that the operators  $\Gamma - a I$  and ${\bar{\Pi}}_0$ have both the rank 1 and are mutually orthogonal. Likewise we conclude from  Eq. (\ref{s9}) that the operators $\Gamma - \tilde{\rho}_j$ and  ${\bar{\Pi}}_j$  are mutually orthogonal   rank-one operators for $j=1,2$. In fact, if for instance $\bar{\Pi}_1$ would be an operator of rank two,  the equality condition in Eq. (\ref{s9}) could be only satisfied when $\Gamma=\tilde{\rho}_1$, which  would violate the positivity constraint $\Gamma-\tilde{\rho}_2 \geq 0$.
We are thus led to the ansatz
\begin{eqnarray}
\label {s11a}
\Gamma \!- a I \!\!&=&\!\! |\tilde{\mu}_0\rangle\langle{\tilde{\mu}}_0|,\;\;\bar{\Pi}_0= Q{|{\bar{\pi}}_0\rangle\langle{\bar{\pi}}_0|},\;\; \langle\tilde{\mu}_0|{\bar{\pi}}_0 \rangle = 0, \;\quad   \\
\label {s11b}
  \Gamma -\tilde{\rho}_j \!\!&=&\! \!|\tilde{\mu}_j\rangle\langle\tilde{ \mu}_j|,\quad  \!\!\bar{\Pi}_j= \beta_j |\tilde{\pi}_j\rangle\langle\tilde{\pi}_j|,\quad \!\!\!\langle\tilde{\mu}_j|{\tilde{\pi}}_j \rangle = 0\;\quad
\end{eqnarray}
 $(j=1,2),$ where we introduced positive constants $\beta_1$ and $\beta_2$ . Here the state vectors characterized by a tilde are non-normalized, while  $|\bar{\pi}_0\rangle$ is normalized to unity in order to yield ${\rm Tr\,}\bar{\Pi}_0= Q$.  
Eqs. (\ref{s11a}) and (\ref{s11b}) imply that for each of the three positive operators $\Gamma - a I$,  $\Gamma -\tilde{\rho}_1$ and $\Gamma -\tilde{\rho}_2$ one of the two eigenvalues is equal to zero. 
Using the orthonormal basis $\{|\nu_1\rangle, |\nu_2\rangle\}$, this requires that  
\begin{eqnarray}
\label {s13a}
|\Gamma_{12}|^2\!\!& =&\!\!(\Gamma_{11}-a)(\Gamma_{22}-a) =(\Gamma_{11}-C_1)(\Gamma_{22}-1+C_2)\nonumber\\
& =&\!\!(\Gamma_{11}-1+C_1)(\Gamma_{22}-C_2),
\end{eqnarray}
due to the fact that the determinants of the three operators have to vanish. With the help of  Eq. (\ref{s13a}) we obtain for each of the three operators the  respective eigenstate that belongs to its zero eigenvalue. Taking into account that  these eigenstates are orthogonal  to the states $|\tilde{\mu}_j\rangle$ $(j=0,1,2)$ and are therefore proportional to $|\bar{{\pi}}_0\rangle$, $|\tilde{\pi}_1\rangle$ and $|\tilde{\pi}_2\rangle$, respectively, we find that  
\begin{eqnarray}
\label {s14a}
|\bar{{\pi}}_0\rangle\!\!& =&\!\!\frac{\sqrt{\Gamma_{22}-a}\, |\nu_1\rangle - e^{-i\delta} \sqrt{\Gamma_{11}-a}\, |\nu_2\rangle}{\sqrt{\Gamma_{11} +\Gamma_{22}-2a}},\\
\label {s14b}
|\tilde{\pi}_1\rangle\! \!&=&\!\!\sqrt{\Gamma_{22}-1+C_2}\, |\nu_1\rangle 
-  e^{-i\delta}\sqrt{\Gamma_{11}-C_1}\, |\nu_2\rangle,\qquad\\
\label {s14c}
|\tilde{\pi}_2\rangle\! \!&=&\!\!\sqrt{\Gamma_{22}-C_2}\, |\nu_1\rangle 
-  e^{-i\delta}\sqrt{\Gamma_{11}-1+C_1}\, |\nu_2\rangle. \qquad
\end{eqnarray}
When  the expressions under the square-root signs are positive and  Eq. (\ref{s13a}) holds,  the positivity conditions in Eqs.  (\ref{s9}) and (\ref{s8}) are satisfied.     

Next we invoke the completeness relation  $\sum_{j=0}^2\Pi_j = I.$ 
Due to  Eq. (\ref{s6a}) the latter relation  takes the form $ \bar{\Pi}_0 +  \bar{\Pi}_1 + \bar{\Pi}_2 =\rho$, leading with  the help of Eqs. (\ref{s11a}) and   (\ref{s11b}) to the  matrix representation  
\begin{equation}
\label {s17}
\rho_{ij}  = Q \langle \nu_i|\bar{\pi}_0\rangle\langle\bar{\pi}_0 |\nu_j\rangle + \sum_{k=1}^{2}\beta_k \langle\nu_i  |\tilde{\pi}_k\rangle\langle\tilde{\pi}_k|  \nu_j\rangle 
 \end{equation}
for $i,j=1,2$. Making use of  Eqs. (\ref{s14a}) -- (\ref{s14c}), we arrive at         
\begin{eqnarray}
\label {s18a}
\rho_{11}+\rho_{22}\!& =& \! 1= Q+ (\beta_1+\beta_2) (\Gamma_{11}+\Gamma_{22}-1)\nonumber\\
&&\quad\!\!\quad+ \;(\beta_1-\beta_2) (C_2-C_1),\\ 
\label {s18b}
\rho_{11}-\rho_{22}\!&=&\! \frac{(\Gamma_{11}+\Gamma_{22})Q}{\Gamma_{11}+\Gamma_{22}-2a} 
+ (\beta_1 +\beta_2) (\Gamma_{22}-\Gamma_{11}) \nonumber\\
&&+(\beta_1 -\beta_2) (C_1+C_2-1),\\
\label {s18c}
-|\rho_{12}|\,e^{i\phi}&\!=\!& \!|\Gamma_{12}|e^{i\delta}\!\left(\!\frac{Q}{\Gamma_{11}+\Gamma_{22}-2a}+\beta_1\!+\beta_2\right)\!.\;\;\;\;\quad
 \end{eqnarray}
Because of Eq.  (\ref{s13a})  we can express  
$\Gamma_{11}$,  $\Gamma_{22}$ and $|\Gamma_{12}|$ as functions of $C_1$, $C_2$ and $a$. Moreover,
the positivity of  $\beta_1 $ and $\beta_2$ requires that $e^{i\delta}= - e^{i\phi}$, as becomes obvious from Eq. (\ref{s18c}).  
Equations (\ref{s18a}) -- (\ref{s18c}) then represent a system of three coupled equations for the unknown parameters $\beta_1$,   $\beta_2$ and  $a$. The solution depends on $Q$. It determines the optimum measurement provided that $Q$ belongs to a certain region where  the expressions under the square-root signs  in Eqs. (\ref{s14a}) --   (\ref{s14c}) and the resulting constants $\beta_1 $ and $\beta_2$ are positive. 
For those values of $Q$ where a solution  fulfilling these conditions  does not exist, 
 we have to search for the optimum measurement anew.  Assuming that one of the two  operators $\bar{\Pi}_1$ or $\bar{\Pi}_2$ vanishes and again supposing that $\Gamma - a I > 0$, we obtain another solution of the optimization problem in dependence of $Q$.   In order to determine its range of validity, we again have to use the positivity constraints.
For the values of $Q$ where also this new  assumption does not yield the optimum measurement we have to drop the  supposition  $\Gamma - a I > 0$, that is, we have to consider the limiting case    $\Gamma - a I = 0$.
Before proceeding, we separately deal with this case.

\subsection{Solution for sufficiently large $Q$}

Let us apply the general treatment of the limiting case    $\Gamma - a I = 0$ to the discrimination of two mixed qubit states. 
First we assume that  $C_2 > C_1$.  As follows from the considerations that led to  Eq. (\ref{lim2}), in this case  the optimality conditions are satisfied when   
 $a=C_2$,  $\bar{\Pi}_1=0$,  and when  $\bar{\Pi}_2=\rho-\bar{\Pi}_0 $ is proportional to $|\nu_{2}\rangle \langle \nu_{2}| $. From $ {\rm Tr\,}\bar{\Pi}_2 = 1-Q$ we then get   
$\bar{\Pi}_2=(1-Q)|\nu_{2}\rangle \langle \nu_{2}|$. The constraint $\bar{\Pi}_0 = \rho -\bar{\Pi}_2  \geq 0$  is fulfilled when  the determinant resulting from the matrix representation of $\bar{\Pi}_0$  is not negative. This yields the requirement $\rho_{11}( \rho_{22}-1+Q) \geq |\rho_{12}|^2$, which holds true for  $Q \geq Q_1$. Here and in the following we use the abbreviations 
\begin{equation}
\label {abb}
 Q_1= \rho_{11}+  \frac{|\rho_{12}|^2}{\rho_{11}}, \quad Q_2= \rho_{22}+  \frac{|\rho_{12}|^2}{\rho_{22}}.
\end{equation}
Since analogous considerations also apply for  $C_1 > C_2$,  Eq. (\ref{lim2}) can be specified,  yielding  the maximum probability of correct results  
\begin{equation}
\label{limit3} 
P_c^{max}\big|_Q \!=\left \{
\begin{array}{ll}
\!\!  C_2(1\!-\!Q) \; & \mbox{for $\; Q\geq Q^{\prime}=Q_1
           $\;\; if  $\,  C_2 \!>\! C_1, \;\, $}  \\
 \!\! C_1(1\!-\!Q) \; & \mbox {for $\; Q\geq Q^{\prime}=Q_2
           $\;\; if  $\,  C_2 \!<\! C_1. \;\, $}
 \end{array}
\right.
\end{equation}
When  $Q=Q^{\prime}$ one of the two eigenvalues of  $\bar{\Pi}_0$ vanishes. Hence the detection operator   ${\Pi}_0$ is a  rank-one  operator, and since the ranks of ${\Pi}_1$ and ${\Pi}_2$ are 0 and 1,  or 1 and 0, respectively,   the measurement is projective.

In the remaining case   $C_1=C_2\equiv C$   we can again apply  the   considerations that led to  Eq. (\ref{lim2}). We then conclude that  the optimality conditions
together with  the constraint   $ \bar{\Pi}_0=\rho-   \bar{\Pi}_1 - \bar{\Pi}_2\geq 0$ 
are satisfied when $a=C$ and when for $j=1,2$   
\begin{equation}
\label{limit4} 
\bar{\Pi}_j = \alpha_j |\nu_j\rangle\langle \nu_j|\quad{\rm with}\;\;(\rho_{11}-\alpha_{1})(\rho_{22}-\alpha_{2})\geq |\rho_{12}|^2.
\end{equation}
Here the two constants  $\alpha_1$ and $\alpha_2$ both have to be nonnegative,   with $\alpha_1+\alpha_2=  {\rm Tr\,}(\bar{\Pi}_1 + \bar{\Pi}_2)=  1-Q$. 
The following cases have to be distinguished:\\
 (i) If   $|\rho_{12}| \leq {\rm min}\{\rho_{11}, \rho_{22}\}$,  the inequality in  
Eq. (\ref{limit4}) holds true whenever  $0\leq  \alpha_j \leq \rho_{jj} -|\rho_{12}|$  $(j=1,2)$, that is whenever the given failure probability falls in the range $2|\rho_{12}| \leq Q \leq 1$. When  $Q=2|\rho_{12}|$  all three detection  operators have the rank 1.\\  
(ii)  If  $|\rho_{12}| \geq {\rm min}\{\rho_{11}, \rho_{22}\}$ we  first assume   that $\rho_{11} <   \rho_{22}$ and therefore $|\rho_{12}| \geq  \rho_{11}$.  Putting $\alpha_1=0$, the constraint expressed in  Eq. (\ref{limit4}) takes the form we discussed before Eq. (\ref{abb}), which is satisfied if $Q \geq Q_1$, where the measurement is projective for $Q=Q_1$.
Similar considerations hold for   $\rho_{22} \leq   \rho_{11}$. 
In summary,  for $C_1=C_2=C$ we obtain   
\begin{eqnarray}
\label{limit5} 
\!\!\!\!&&P_c^{max}\big|_Q\!= \! C(1-Q)\;\;\; {\rm if}\;\; Q\geq Q^{\prime}=Q_{min}^{MC}  \\
\label{limit6} 
\!\!\!\!&&{\rm with}\;\;Q^{\prime}=\left \{
\begin{array}{ll} 
 2 |\rho_{12}|\;  & \mbox
{if $ \;|\rho_{12}| \leq {\rm min}\{\rho_{11}, \rho_{22}\},$}\qquad\\
 Q_1   & \mbox{if $\;
            |\rho_{12}| \geq \rho_{11},$}  \quad\\
  Q_2   &\mbox{if $\;
           |\rho_{12}|\geq  \rho_{22}. $}  \quad
 \end{array}
\right.
\end{eqnarray}
In Eq. (\ref{limit5}) we took into account that for $C_1=C_2$  the value  of $Q^{\prime}$ determines  the minimum failure probability $Q_{min}^{MC}$  necessary for maximum-confidence discrimination,  see Eq. (\ref{conf1}). We remark that in our earlier paper \cite{herzog1} we calculated  $Q_{min}^{MC}$ for two  mixed qubit states and found that the latter is given by the expressions in  Eq. ( \ref{limit6})  for arbitrary values of  $C_1$ and $C_2$. 

As outlined in Sec. II after Eq. (\ref{rel-rate}), a discrimination measurement   maximizing  $P_c$  when $Q$ is fixed at a value  $Q > Q^{\prime}$ is without practical importance. In the present case such a measurement would be  a generalized measurement realizable in an extended Hilbert space, where projections onto two orthogonal directions would indicate an inconclusive result since $\bar{\Pi}_0$ has the rank 2,  which implies that also $\Pi_0$ is a  rank-two operator.

\subsection{Complete solution for the case  ${\bf  C_1=C_2\equiv C}$} 

In order to treat the case $\Gamma-a I>0$  we use the method developed in Sec. III A. It turns out that for $C_1 \neq C_2$ this leads to a fourth order polynomial equation in the variable $a(1-a)$. To obtain a simple analytical solution  we therefore assume that   $C_1=C_2\equiv C$, restricting ourselves to cases where  the  maximum achievable confidence is equal for the two possible outcomes. 
Here we present  the final result for the complete solution,   
while the details of the derivation are shown in Appendix A. 
 Two cases have to be considered: \\
\noindent  (i) If  $|\rho_{12}| \leq {\rm min}\{\rho_{11}, \rho_{22}\}$ we find that the maximum probability  of correct results with   arbitrary values of the fixed failure probability  $Q$ is given by       
\begin{eqnarray}
\label{c6} 
&&P_c^{max}\big|_Q \!=\left \{
\begin{array}{ll}\!\!  P_c ^{(0)}(Q)  \;\; & \mbox{if $\;
            0\leq Q\leq 2|\rho_{12}|   $},  \qquad\\
 \!\! C(1-Q)\; \; & \mbox{if $\;
               2|\rho_{12}|\leq Q  \leq 1 $},
 \end{array}
\right.
\end{eqnarray}
where 
\begin{equation}
\label {c7}
P_c ^{(0)}\!= \!\frac{1-Q}{2}+\frac{2C-1}{2}\sqrt{ ({1} \!-\!2|\rho_{12}|)\, ({1}\!+\! 2|\rho_{12}|\!-\!2Q)}.
\end{equation}
\noindent (ii)  If  $|\rho_{12}| \geq {\rm min}\{\rho_{11}, \rho_{22}\}$ we assume without lack of generality that  $\rho_{11} < \rho_{22}$,  which means that $|\rho_{12}| \geq  \rho_{11}$. Then  we get  
\begin{eqnarray}
\label{c8} 
&&P_c^{max}\big|_Q \!=\left \{
\begin{array}{ll}\!\!  P_c ^{(0)}(Q)  \;\; & \mbox{if $\;
          \; \;0\;\,\leq Q\leq Q_ {cr}   $},  \qquad\\
 \!\!P_c ^{(1)}(Q) \; \; & \mbox{if $\;
               Q_ {cr}\leq Q  \leq Q_1 $},\qquad\\
\!\! C^{}(1-Q)\; \; & \mbox{if $\;
               Q_1\;\leq Q \leq \;1, $}\
 \end{array}
\right.
\end{eqnarray}
where $Q_1$  is defined in  Eq. (\ref{abb}). $Q_ {cr}$ is given by
\begin{equation}
\label {c9}
Q_ {cr}=\frac{2\Delta}{1-2|\rho_{12}|}\quad {\rm with}\;\; \Delta= \rho_{11}\rho_{22} - |\rho_{12}|^2
\end{equation}
and denotes a critical failure probability that separates the regions where the optimum measurement is a generalized measurement and where it is projective, see below. 
Note that  $\Delta > 0$ since $\rho=\eta_1\rho_1+ \eta_2\rho_2$ is a mixed state.   In Eq. (\ref{c8}) we introduced 
\begin{eqnarray}
\label {c10}
 P_c^{(1)}\!&\!=\!& C(1\!-\!Q) - \frac{2C\!-\!1}{1\!-\!4\Delta} \Big[\Delta (1\!-\!2\rho_{11})\\
&&\quad -(2\Delta \!-\! \rho_{11})(1\!-\!Q)-2|\rho_{12}|\sqrt{\Delta(Q\!- \!Q^2 \!- \!\Delta) }\Big].\nonumber
\end{eqnarray}
 The  corresponding result for $ \rho_{22} < \rho_{11}$ 
 is obtained when in  Eqs.  (\ref{c8}) and  (\ref{c10})  $Q_1$ and $\rho_{11}$ are replaced  by $Q_2$ and $\rho_{22}$, respectively. 

When the conditions  in the upper lines of Eqs. (\ref{c6}) or (\ref{c8}), respectively,  are fulfilled,  the optimum detection operators are given by Eq. (\ref{s24c}) together with Eq. (\ref{s24}) and Eqs. (\ref{s24d}) --  (\ref{s24f}), see Appendix A. They describe a generalized measurement if $ Q \neq 0$ and $Q \neq  Q_ {cr}$.  For $Q=0$ the optimum measurement is equal to the projective measurement for minimum-error discrimination,  
yielding the maximum probability of correct results  $P_c^{max}\big|_{Q=0}= 
P_c^{ME}=\frac{1}{2}(1+{\rm Tr\,}\|\eta_2\rho_2-\eta_1\rho_1\|)$ \cite{helstrom,holevo}.

When the middle line of Eq. (\ref{c8}) applies,  the optimum detection operators follow from Eq. (\ref{s28}) together with Eqs. (\ref{s28a}) and  (\ref{s32}). In this case the optimum measurement is a projective measurement, where only one of the states, here the state $\rho_2$,  is conclusively discriminated with a certain probability, while in the presence of the other state  an inconclusive result is always obtained.  
The lower lines of Eqs. (\ref{c6}) or (\ref{c8}), respectively, correspond  to the limiting case  $Q\geq Q^{\prime}$ discussed in the previous paragraph. 

In the special case when $C=1$, $ \rho_{11}=\eta_1$,  $ \rho_{22}=\eta_2$, and $ \rho_{12}=\sqrt{\eta_1\eta_2}\,\langle\psi_1|\psi_2\rangle$  our solution refers to the discrimination of two pure states, see   Eq. (\ref{s11}).  Two arbitrary qubit states that are mixed can be  represented as    
\begin{equation}
\label {c1} \rho_{1}\!=\! p_1|\psi_1\rangle\langle \psi_1|+\!\frac{1\!-\!p_1}{2}\,I, \quad
 \rho_{2}\!=\! p_2|\psi_2\rangle\langle \psi_2|+\!\frac{1\!-\!p_2}{2}\,I, 
\end{equation}
where  $|\psi_1\rangle$  and $|\psi_2\rangle$ are normalized pure states with the overlap $\langle \psi_1|\psi_2 \rangle \equiv S$ and where ${\rm Tr\,}I=2$.  The parameters $p_1$ and $p_2$  with $0\leq p_1, p_2  \leq 1$ are related to the purities of the two states.  
The condition 
$C_1=C_2=C$ with $C<1$ is fulfilled provided that $\left(\eta_1^{-1}-1\right)^2=\frac{1-p_1^2}{1-p_2^2}$, see Appendix A. Clearly,  for any two mixed qubit states $\rho_1$ and $\rho_2$  there exist particular prior probabilities of occurrence for which our solution holds, and  for any given prior probability $\eta_1$ we can find a whole class of state pairs where the solution applies.  
In Figs. 1 and 2 the ratio $R_c^{max}|_Q= P_c^{max}|_Q/(1-Q)$ is plotted for the discrimination of pure states ($C=1$) and of mixed states ($C<1$). The smallest value of $Q$ where $R_c^{max}|_Q=C$  is equal to the value $Q^{\prime}$ given by  Eq. (\ref{limit6}) and corresponds to  the minimum failure probability  necessary for unambiguous discrimination if $C=1$ 
   \cite{jaeger},  or necessary for  maximum-confidence discrimination if $C<1$  \cite{herzog1}.  
\begin{figure}[t!]
\center{\includegraphics[width=7 cm]{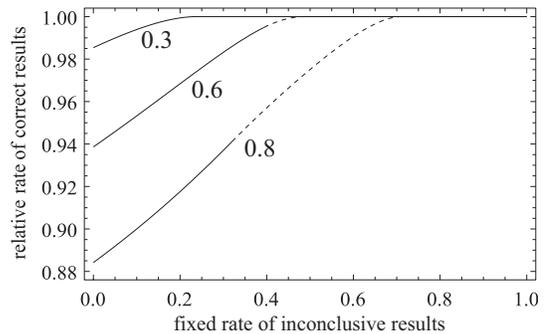} }
\caption{Maximum relative 
rate of correct results $R_c^{max}|_Q = P_c^{max}|_Q/(1-Q)$ versus  the fixed rate of inconclusive results $Q$ for discriminating two pure states with $|S|= | \langle \psi_1|\psi_2 \rangle|= 0.3$, 0.6, and 0.8,  occurring with  the prior probabilities  $\eta_1= 0.2$ and $\eta_2 = 1- \eta_1$.   The dashed lines refer to the regions where the optimum measurement is projective,  detecting only the second state (see text). }
\end{figure}
\begin{figure}[t!]
\center{\includegraphics[width=7 cm]{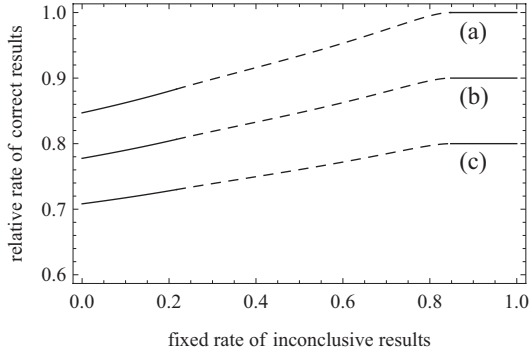} }
\caption{Same as Fig. 1 but including mixed qubit states, where $C< 1$. The parameters are  $\rho_{11}=0.2$,   $|\rho_{12}|=0.36,$ and (a) $C=1$,  (b) $C=0.9,$  and (c) $C=0.8$. When the states are represented by  Eq.  (\ref{c1}),     this corresponds to the cases that (a) \{$\eta_1= 0.2$,  $p_1=p_2=1$, $ |S|=0.9$\}, (b) \{$\eta_1=0.26, \;p_1=0.92,\;p_2=0.99,\;|S|=0.95$\}, and (c) \{$\eta_1= 0.32,\;p_1=0.90,\;
p_2=0.98,\;|S|=0.96$\}. }
\end{figure}

We  emphasize that for $Q=0$ and for sufficiently large values of $Q$  the analytical solution of our optimization problem is known for two arbitrary qubit states, occurring with arbitrary prior  probabilities.
The first case corresponds to minimum-error discrimination \cite{helstrom,holevo},  where for two states the measurement is projective and  can be easily determined analytically when the Hilbert space is two-dimensional. The second case was treated  in Sec. III B  and applies when $Q\geq Q^{\prime}$, where $Q^{\prime}$ is given by   Eqs. (\ref{limit3}) or  (\ref{limit6}), respectively.  For $0<Q<Q^{\prime}$, however, we obtained a simple analytical solution  only for those states and prior probabilities where  the maximum confidence is equal for the two outcomes, $C_1=C_2\equiv C$.   

 \section{$N$ symmetric states}

\subsection{Optimality conditions  for equiprobable states and the limiting case of sufficiently large $Q$} 

Another class of analytically solvable cases refers to discriminating $N$ states 
that are symmetric, which means that for $j=1,\ldots,N$  
\begin{equation}
\label {t1} \rho_{j}= V^{(j-1)}\rho_1 V^{\dag( j-1) } \quad {\rm with}
\quad V^{\dag}V= V^{N}=I
\end{equation}
 \cite{ban}, where without lack of generality an arbitrary state of the given set of states can be chosen as the reference state $\rho_1$. 

Assuming that each state is prepared with the same prior probability, $\eta_j = 1/N$, we find that  $\rho=\frac{1}{N}\sum_{j=1}^N \rho_j =V \rho V^{\dag}$
and thus $[V,\rho]=0$, which implies that the Hermitian operator $\rho$ and the unitary operator $V$ can be diagonalized in the same orthonormal basis \cite{ban}.  Introducing the orthonormal basis states $|r_l\rangle$, we get the spectral representations 
\begin{equation}
\label{t2} 
\rho\!= \!\frac{1}{N}\sum_{j=1}^N \rho_j \!=\! \sum_{l=1}^d r_l|r_l\rangle\langle r_l|, \quad V\!= \! \sum_{l=1}^d v_l|r_l\rangle\langle r_l|,
\end{equation}
 where $\sum_{j=1}^{N} (v_lv_{l^{\prime}}^{\ast})^{j} = N\delta_{ll^{\prime}}$ and 
 $|v_l|^2=v_l^N=1$,
 see also  \cite{herzog2}. 
We now  focus on the optimum measurement, where the optimality conditions, given by Eqs. (\ref{Q2}) and (\ref{Q4}), are satisfied. 
Let us suppose that $\Pi_1$ is an element of the set of optimum detection
operators. We  introduce   
\begin{equation}
\label {t3} \Pi_{j}= V^{(j-1)} \Pi_1 V^{\dag (j-1)}
\end{equation}
 ($j=1,\ldots, N$), where due to the symmetry  the operator
$\sum_{j=1}^N\Pi_j =I-\Pi_0$ commutes with $V$. Using $[V,\rho]=0$, this  implies that 
\begin{equation}
\label {s7c}  [\Pi_0,V]=  [\Pi_0,\rho]= [\Pi_0,Z] =[\rho,Z]=[V,Z]=0.
\end{equation}
Here the second equality sign follows from the fact that in the optimum measurement
 $(Z-a\rho)\Pi_0= \Pi_0(Z-a\rho)=0,$ which leads to  $[\Pi_0,Z]=a[\Pi_0,\rho]$. 
Upon inserting Eq. (\ref{t3}) into  Eqs. (\ref{Q2}) and (\ref{Q4}) with $\eta_j=1/N$,  taking Eq. (\ref{s7c}) into account, 
 it becomes obvious that if  $\Pi_1$ fulfills the optimality conditions,  then these conditions
are also fulfilled by each of the operators $\Pi_j$ defined in Eq. (\ref{t3}), in analogy to the derivation of the optimality conditions for maximum-confidence discrimination of symmetric mixed states in our previous paper \cite{herzog2}.  
Hence  the detection operators for the optimum measurement
 can always be chosen in the form of Eq. (\ref{t3}). 
The optimality conditions  thus reduce to  the conditions
\begin{eqnarray}
\label {opt-symm1}
\,Z-a\rho &\geq& 0,\quad (Z-a\rho)\;\Pi_0 = 0,\\
\label {opt-symm2}
 Z-\frac{\rho_1}{N} &\geq& 0,\quad  \left(\!Z-\frac{\rho_1}{N}\right)\!\Pi_1=0, 
\end{eqnarray}
which were first derived by applying group-theoretical methods \cite{eldar1}.
Using   Eq. (\ref{t3}) and the properties of $V$ we find that    
\begin{equation}
\label {opt-symm3} 
\Pi_0= \sum_{l=1}^{d}\big (1-N\langle r_l|\Pi_1|r_l\rangle\big )|r_l\rangle\langle r_l|,\quad Z=\sum_{l=1}^d z_l |r_l\rangle\langle r_l|,
\end{equation}
where we introduced the spectral representation of $Z$, taking into account that  $\Pi_0$ and $Z$ commute when the optimality conditions are fulfilled. With the help of  Eqs. (\ref{Qdual5}) and (\ref{t3})  we can represent   the maximum  probability of correct results  
achievable with the fixed value  $Q= {\rm Tr\,} (\rho\Pi_0)$ as  
\begin{equation}
\label {opt-symm3a}
P_c^{max}|_Q = {\rm Tr\,}(\rho_1\Pi_1 )\quad{\rm where}\;\; 
Q=1-N {\rm Tr\,}(\rho\Pi_1 ). 
\end{equation}

Let us  discuss the limiting case of  sufficiently large $Q,$ which we treated in general in Sec. II B. 
According to Eq. (\ref{limit0})  
in this limit the optimality conditions,  Eqs. (\ref{opt-symm1}) and  (\ref{opt-symm2}), reduce to the conditions   $ aI-\tilde{\rho}_1\geq 0$ and  $\left(aI-\tilde{\rho}_1\right)\bar{\Pi}_1=0,$ where we used the transformed operators  introduced in Eqs. (\ref {s6a}) and  (\ref {s6}).  
 The spectral representation of  $\tilde{\rho}_1$ can be written as 
\begin{equation}
\label  {opt-symm3b}
\tilde{\rho}_1 
=\frac{1}{N}\rho^{-1/2}  \rho_1  \rho^{-1/2}=C \sum_{k=1}^{k_1}  |\nu_k\rangle\langle
\nu_k| +\!\! \sum_{k=k_1+1}^d
\!\!\!\nu_k|\nu_k\rangle\langle \nu_k|
\end{equation}
 where $C$ denotes the largest eigenvalue and $k_1$ is its degree of degeneracy. 
The optimality conditions are fulfilled when $a=C$ and when  
the support of $\bar{\Pi}_1 = \rho^{1/2}\Pi_1  \rho^{1/2}$ is the Hilbert space spanned by the eigenstates $ |\nu_1\rangle \ldots  |\nu_{k_1}\rangle$. We then obtain   
$P_c^{max}\big|_Q =
C N {\rm Tr\,}\bar{\Pi}_1 $, which yields
\begin{equation}
\label  {opt-symm3e}
P_c^{max}\big|_Q =
 C(1-Q) \quad {\rm if} \;\;Q\geq Q^{\prime}=Q_{min}^{MC}. 
\end{equation}
Here we took into account that the maximum confidence is the same for discriminating each of the equiprobable symmetric states 
${\rho}_j = V^{(j-1)}\rho_1 V^{\dag( j-1) }$ since the largest eigenvalues of the operators $\tilde{\rho}_j$  are identical, that is,  
\begin{equation}
\label {opt-symm3f}
 C_1=\dots = C_N\equiv C.
\end{equation}
    $Q^{\prime}$  therefore corresponds to the failure probability $Q_{min}^{MC}$ in optimized  maximum-confidence discrimination, or,  if $C=1$, in optimum  unambiguous discrimination, respectively,  see Eq. (\ref{conf1}).  
In order to determine  $Q^{\prime}$ we have to find the minimum of $Q=1-N{\rm Tr\,}\bar{\Pi}_1$ on the condition that  with the given support of $\bar{\Pi}_1$  the positivity constraint  $\bar{\Pi}_0 \geq 0$ is satisfied, which because of  Eq. (\ref{opt-symm3})  is equivalent to the constraint  
$ \sum_{l=1}^{d}\left[r_l-N\langle r_l|\bar{\Pi}_1|r_l\rangle\right] |r_l\rangle\langle r_l|\geq 0$, see  \cite{herzog2}.   
In general, this is a nontrivial task. In the special case when the largest eigenvalue of $\tilde{\rho}_1$ is nondegenerate, $\bar{\Pi}_1$ is  proportional to $|\nu_1\rangle\langle \nu_1|$.  In agreement with our previous paper \cite{herzog2} we then  obtain      
$Q^{\prime}=Q_{min}^{MC} = 1- \underset{\overset{l}{}}{\mathrm{min} } \left \{\frac{r_l}{ |\langle r_l|\nu_1\rangle|^2}\right \}$
for $k_1=1$.

\subsection{$N$ equiprobable  symmetric pure qudit states}

\subsubsection{Solution for a special class of states}

First we consider the discrimination  of  $N$ symmetric equiprobable  pure states   spanning a $d$-dimensional Hilbert space  ${\cal H}_d$. This means  that $d \leq N$ and  implies that all expansion coefficients $c_l$ of  the states with respect to the $d$-dimensional eigenbasis of the symmetry operator $V$ are different from zero \cite{herzog2}. In  general, $N$ symmetric pure qudit states  are given by 
\begin{equation}
\label {sol-0} 
 |\psi_{j}\rangle = V^{(j-1)}|\psi_1\rangle \;\; {\rm with} \;\;   |\psi_1 \rangle =  \!\sum_{l=1} ^d  c_l  |r_l\rangle \;\;( j=1,\ldots,N).
\end{equation}
In order to obtain an analytical solution  we restrict ourselves to special states where only two different values of the expansion coefficients $c_l$ occur. We assume that   
\begin{equation}
\label {sol-1}    |\psi_1 \rangle \!=\! c_1 \!\sum_{l=1}^m  |r_l\rangle \!+\!  c_2 \!\! \!\sum_{l=m+1}^d\! \!\! |r_l \rangle,\quad m|c_1|^2+(d\!-\!m)|c_2|^2\!=\!1,
\end{equation}
where  $m\leq d/2$ and where we distinguish the cases   $|c_1|\geq |c_2|$ and $|c_1|\leq |c_2|$ with $c_1,c_2 \neq  0$.   In  Appendix B 
we derive the maximum probability of correct results $P_c^{max}\big|_Q$  with the fixed probability $Q$ of inconclusive results,  making use of the optimality conditions, Eqs. (\ref{opt-symm1}) and  (\ref{opt-symm2}).     
Using Eq. (\ref{opt-symm12}) with $p=1$ and Eq. (\ref{opt-symm20})   we obtain   
\begin{equation}
\label{sol-2} 
P_c^{max}\big|_Q \!=\left \{
\begin{array}{ll}\!\!  P_c ^{(0)}(Q) & \mbox{if\, $
            Q\leq Q^{\prime} =\!1\!-\!d\, {\rm min}\{|c_1|^2,|c_2|^2 \}  $} \\
\!\! \frac{d}{N}(1-Q) & \mbox{if \,$
            Q \geq Q^{\prime} $}
 \end{array}
\right.
\end{equation}
with 
\begin{equation}
\label{sol-3} 
\!P_c^{(0)} \!=\!\left \{
\begin{array}{ll}\!\!\!  \frac{1}{N}\!\left[ \sqrt{m}\sqrt{m|c_1|^2-Q}+ (d-m)|c_2|\right]^2 \;\; \;\; \mbox{if $\!
       \frac{  | c_1|}{|c_2|} \!\geq\! 1     $}  \qquad\\
\!\!\! \frac{1}{N}\!\left[\sqrt{d\!-\!m}\sqrt{(d\!-\!m)|c_2|^2-Q}
+m|c_1|\right]^2  \mbox{if $
       \!  \frac{  | c_1|}{|c_2|} \!\leq\! 1   $}.  
\end{array}
\right.
\end{equation}
The lower line of Eq. (\ref{sol-2}) follows from Eq. (\ref {opt-symm3e}) and from the fact  that $P_c^{(0)}(Q^{\prime})/(1- Q^{\prime})
=C=\frac{d}{N}$.  The  optimum detection operators are given by   Eqs.  (\ref{y0}) and (\ref{opt-symm7}). 

When $Q=0$ the optimum measurement  corresponds to  minimum-error discrimination. For this case the optimum  detection operators discriminating the equiprobable states given by Eq. (\ref{sol-0}) are  known to be   $\Pi_{j}^{ME}=
\frac{1}{N}\rho^{-1/2} |\psi_j\rangle\langle \psi_j |\rho^{-1/2}$ \cite{ban}, yielding for an arbitrary reference  state $ |\psi_1 \rangle $
the maximum probability of correct results  $P_c^{max}\big|_{Q=0}=\frac{1}{N}\left(\sum_{l=1}^d| c_l|\right)^2$. 
 On the other hand, when  $Q= Q^{\prime}=Q_{min}^{MC}$  the optimum measurement  
corresponds to optimized maximum-confidence discrimination. For $N$ equiprobable symmetric pure qudit states with  an arbitrary reference state $ |\psi_1 \rangle$  it has been shown that  $ Q_{min}^{MC} = 1-d \,{\rm
min}_l \{|c_l|^2\}$  \cite{herzog2,jimenez}, including the special case of unambiguous discrimination  when $N=d$ \cite{chefles-symm}. 
The results obtained from Eq. (\ref{sol-2}) for  the maximum confidence $C$, for $Q^{\prime}
= Q_{min}^{MC}$, and for  $P_c^{max}\big|_{Q=0}$ agree with these  previous results.  
Note that for  $|c_1|=|c_2|=\frac{1}{d}$ we obtain $Q^{\prime}=0$,  which means that in this case minimum-error discrimination and optimized maximum-confidence discrimination are equivalent, cf. also Sec. V A.  

We emphasize that while in the limiting cases  $Q=0$ and  $Q=Q^{\prime}$ the optimum measurement with a fixed value of $Q$  
is known for $N$ equiprobable  symmetric pure  qudit states which are arbitrary, 
our complete analytical solution, Eq. (\ref{sol-2}), that interpolates between these limiting cases, is restricted to the special class of those symmetric states  where  the reference state is given by Eq. (\ref{sol-1}).

\subsubsection{Application to  $N$ linearly  independent  symmetric states}

 When the number of pure states is equal to the dimension of the Hilbert space spanned by them,  $N=d$, the states are linearly independent. This means that without lack of generality the  eigenvalues of the symmetry operator $V$ in Eq. (\ref{t2}) can be written as  $ v_l={\rm exp}\left({2\pi i
\frac{l-1}{N}}\right)$ for $l=1, \ldots, N$  \cite{chefles-symm}. With the help of  Eqs. (\ref{sol-0}) and (\ref{sol-1}) we obtain for $d=N$ the mutual overlaps $\langle \psi_j|\psi_k\rangle= \sum_{l=1}^N v_l^k v_l^{*j} |\langle \psi_1|r_l\rangle|^2$, yielding with  $j\neq k$ 
\begin{equation}
\label {sol-4}  
\langle \psi_j|\psi_k\rangle= (|c_1|^2 -  |c_2|^2)\sum_{l=1}^m {\rm exp}\mbox{$ \left[2\pi i 
\frac{l-1}{N}(k\!-\!j)\right]$}.
\end{equation}
Clearly, when  $m=1$ all mutual overlaps are equal and real. If $d=N=3$ this  is always the case, because of the requirement  $m\leq d/2$ in Eq. (\ref{sol-1}).    For $d=N\geq 4$ and $m \geq 2 $, however,  different values of the overlap  occur for the linearly independent symmetric states specified by Eq. (\ref{sol-1}).  
  
We now specialize to  $N$  linearly independent symmetric pure states with equal mutual overlaps $S$,  where   
\begin{equation}
\label {sol-4a}  
 \langle\psi_j|\psi_k\rangle  \equiv S=|c_1|^2 -  |c_2|^2= 1-N |c_2|^2 
\end{equation}
according to Eq. (\ref{sol-4}) with $m=1$.       
 Using  the normalization condition given in Eq. (\ref{sol-1}), we get  the expressions  
$ |c_1|^2 = \frac{1+(N-1)S}{N}$ and $|c_2|^2=\frac{1-S}{N}$, from which we conclude  that 
$-(N\!-\!1)^{-1}\! \leq  S \leq 1\!$ since neither  $|c_1|^2$ nor  $|c_2|^2$ can be negative. 
Upon inserting these expressions into  Eq. (\ref{sol-3}) we obtain for $|c_1| \geq |c_2|$, that is for $S\geq 0$,  
\begin{eqnarray}
\label {sol-5a}  
\!\!\!&&\!\!\!\!\!P_c^{(0)}=\frac{1}{N}\mbox {$\left[\sqrt{\frac{1+(N-1)S}{N}-Q}+(N-1)\sqrt{\frac{1-S}{N}}\right]^2$}\quad\nonumber \\
\!\!\!&&\!\!\!\!\! \mbox{if \;\;$0 \leq  S  < 1,$ \,\;\qquad  for \;\;
$ Q \leq Q^{\prime}= S,  $ \;} 
\end{eqnarray}
while for negative overlaps we arrive at  
\begin{eqnarray}
\label {sol-5b}  
&&\!\!\!P_c^{(0)}\!=\!\frac{1}{N}\mbox {$\left[\sqrt{N\!-\!1}\sqrt{\frac{(N-1)(1-S)}{N}-Q}
+\sqrt{\frac{1+(N-1)S}{N}}\right]^2$}\nonumber \\
&&\!\!\!\mbox{if $\;\;- \frac{1}{N-1}\! <  S \leq 0,\!$ \quad  for \;
$ Q \leq Q^{\prime}\!=\! (N\!-\!1) |S|.$ } 
\end{eqnarray}
If  $Q\geq Q^{\prime}$ Eq. (\ref{sol-2}) yields   $P_c^{max}|_Q=1-Q$, which corresponds to unambiguous discrimination since the probability of errors vanishes in this case, $P_e=1-Q-P_c=0$. The minimum failure probability necessary for unambiguous discrimination is thus given by  $Q^{\prime}$.
When $S=0$ Eqs. (\ref{sol-5a}) and   (\ref{sol-5b}) apply only for $Q=0$ and we get  
$P_c^{max}\big|_{Q=0}=1$, as expected for mutually orthogonal states. On the other hand,  in the limit  where $S$ approaches $- \frac{1}{N-1}$ and     $c_1$ therefore approaches zero, 
the states get linearly dependent and span a Hilbert space of dimension $N-1$ since $m=1$. 
For example, if $N=3$  and $  \langle \psi_j|\psi_k\rangle =  -\frac{1}{2}=\cos \frac{2\pi}{3}$ $(j\!\neq \!k)$ we arrive at the  trine states that can be represented by  three  real symmetric state vectors spanning a two dimensional  Hilbert space.
We note that  for $N=2$ Eqs. (\ref{sol-5a}) and (\ref{sol-5b}) are identical and reproduce the result obtained for the discrimination of two equiprobable pure states \cite{chefles-barnett}. 
For $N\geq 3$, however,  the explicit expressions for $P_c^{max}$ and $Q^{\prime}$ depend on the sign of the overlap between the states, see Fig. 3.   
\begin{figure}[t!]
\center{\includegraphics[width=7 cm]{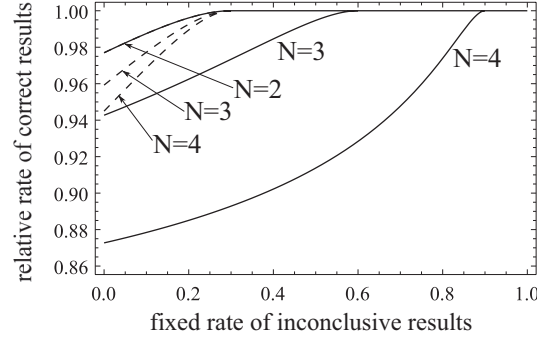}} 
\caption{Maximum relative 
rate of correct results $R_c^{max}|_Q = P_c^{max}|_Q/(1-Q)$  versus the fixed rate of inconclusive results $Q$ for discriminating $N$ equiprobable linearly independent  symmetric pure states with equal mutual overlaps  $S=-0.3$ (full lines) and $S=0.3$ (dashed lines). For $N=2$ both lines coincide.}
\end{figure}
\begin{figure}[t!]
\center{\includegraphics[width=7 cm]{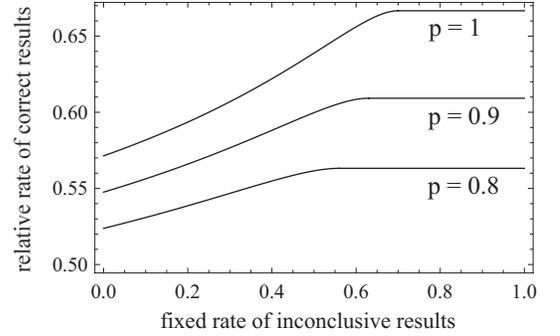} }
\caption{Same as Fig. 3   for discriminating $N = 3$ equiprobable symmetric mixed qubit states, characterized by Eq. (\ref{sol-6}) with  $|c_1|^2 = 0.85$ and $|c_2|^2 = 0.15 $,  for different values of the purity parameter $p$. }
\end{figure}

\subsection {$N$ equiprobable symmetric mixed  qubit states}

The density operator of any mixed qubit state can be written as $p\,|\psi\rangle\langle \psi|+
\frac{1-p}{2}\,I $ with a certain normalized pure state $|\psi\rangle$ and with $0 <p<1$. Hence 
 the most general representation  for  $N$ symmetric mixed states in a two-dimensional joint Hilbert with the identity operator $I$  is given by  the density operators  $ \rho_{j} = V^{(j-1)}\rho_1   V^{\dag(j-1)}$  with  $j=1,\ldots,N$ and with $ V^{\dag}V= V^{N}=I,$
where 
\begin{equation}
\label {sol-6}
 \rho_{1} = p\,|\psi_1\rangle\langle \psi_1|+
\frac{1-p}{2}\,I, \quad     |\psi_{1}\rangle \!\!= c_1|r_1\rangle + c_2|r_2\rangle.                 
\end{equation}
Without lack of generality we assume that  $|c_1|\geq |c_2| $.  
 Using  Eqs.  (\ref{opt-symm8}), (\ref{opt-symm12})  and  (\ref{opt-symm21}) (see  Appendix B) we find that  the maximum probability of correct results  with  the fixed failure probability $Q$ is given by
\begin{eqnarray}
\label{sol-7} 
P_c^{max}\big|_Q \!=\left \{
\begin{array}{ll}\!\!  P_c ^{(0)}(Q)  \;\; & \mbox{if $\;
              Q\leq Q^{\prime}=   p (1 -2|c_2|^2)  $}  \quad\\
\!\! C(1-Q)\; \; & \mbox{if $\;
            Q\geq Q^{\prime} $}\
 \end{array}
\right.
\end{eqnarray}
\begin{eqnarray}
\label {sol-8}  
{\rm with\;}\;\;P_c^{(0)} & =& \frac{p}{N}  {\left [ |c_1|\sqrt{1-\frac{2Q}{1+p(1-2|c_2|^2)}}+|c_2| \right]^2}\nonumber \\
&&+ \frac{1-p}{N}{\left [1-\frac{Q}{1+p(1-2|c_2|^2)}\right]}. 
\end{eqnarray}
The ratio $P_c^{max}|_Q/(1-Q)$  is plotted in  Fig. 4. By calculating  $ C =   P_c^{(0)}(Q^{\prime})/(1- Q^{\prime})$ we arrive at the maximum confidence  $C =  \frac{1}{N}\left(1+\frac{2p\,|c_1
c_2|}{\sqrt{1-p^2(1-2|c_2|^2)^2} }\right)$,
in accordance with the result derived for $C$ in our previous paper \cite{herzog2}. 
  $Q^{\prime}$ coincides with the smallest failure probability necessary to achieve  maximum-confidence discrimination   \cite{herzog2}. 
On the other hand, for $Q=0$ we get the result 
\begin{eqnarray}
\label {sol-8a}  
P_c^{max}\big|_{Q=0} = P_c^{ME}\!= \frac{1}{N}\left(1+p|c_1c_2|\right), 
\end{eqnarray}
which means that using our general solution for $P_c^{max}\big|_{Q}$ we determined the maximum probability of correct results in  minimum-error discrimination of the $N$ mixed qubit states.  

 The special case $|c_1|^2= |c_2|^2=0.5$ is worth mentioning,  where  $Q^{\prime}=0$ and therefore $P_c^{max}/(1-Q) =C=\frac{1+p}{N}$ for any value of $Q$.   The measurements for minimum-error discrimination and for optimized maximum-confidence discrimination are then the same, cf. Sec. V A. When  $N=3$ this  applies for the depolarized trine states, described by  Eq. (\ref{sol-6}) with  $|\psi_1 \rangle = \frac{1}{\sqrt{2}}\left [{\rm exp}\left(i\frac{2\pi}{3}\right)|r_1\rangle +  {\rm exp}\left(-i\frac{2\pi}{3}\right)|r_2\rangle \right]$.

\subsection{$N$ special  symmetric mixed states of rank $D$ spanning a joint Hilbert space of dimension $ND$} 

In the following we treat a special case where the optimum measurement for discriminating $N$ symmetric mixed states with a fixed probability $Q$ of inconclusive results can be obtained by applying the pure-state solution. 
We consider $N$ mixed states of rank $D$, occurring with the prior probabilities $\eta_j$ $(j=1,\ldots,N)$ and being  
described by the special  density operators   
\begin{equation}
\label {ND1}
 \rho_j\!=\!  \sum_{l=1}^D\!  s_{ll^{\prime}} |s_j^{l}\rangle \langle s_j^{l^{\prime}}|\quad {\rm with}\quad 
\langle s_j^{l}|  s_j^{l^{\prime}}\rangle =  \delta_{ll^{\prime}},
 \end{equation}
where we assume that  the overlaps of basis states belonging to different density operators obey  the special relation   
\begin{equation}
\label{ND2} 
\langle s_i^{l}|  s_j^{l^{\prime}}\rangle  =
  S\,\delta_{ll^{\prime}}\quad 
\mbox{with\; $  - \frac{1}{N-1}< S <  1$} \quad (i\neq j)    
\end{equation}
and with  $l, l^{\prime}=1,\ldots,D$. 
For each value  of $l$ the states $|s_1^l\rangle, \ldots,|s_N^l \rangle $ form a set of $N$ linearly independent symmetric states with equal mutual overlaps, as becomes obvious from Eq. (\ref{sol-4a}).   
The spectral representations of the density operators can be written as  
\begin{equation}
\label {ND3}
 \rho_j \!=\!\sum_{k=1}^D  \lambda_k \,\rho_j^{(k)}\quad  {\rm with}\quad
\rho_j^{(k)}\!= |\psi_j^{k}\rangle \langle \psi_j^{k}|,\quad \sum_{k=1}^D \lambda_k\!=\!1,
 \end{equation}
where  $\langle \psi_j^{k}|  \psi_j^{k^{\prime}}\rangle \! = \!
 \delta_{kk^{\prime}}$ and   $\langle \psi_i^{k}|  \psi_j^{k^{\prime}}\rangle \! = \!
  S\,\delta_{kk^{\prime}}$ for  $i\neq j$. The latter equation  follows  from   Eq. (\ref{ND2}) after expanding the normalized eigenstates as  $ |\psi_j^{k}\rangle = \sum_{l} u_{kl} |s_j^{l}\rangle$,   taking into account that the expansion coefficients $u_{kl}=\langle  \psi_j^{k}| s_j^{l}\rangle $ and the eigenvalues $\lambda_k$ are identical for the different states $j$  since  the matrix elements $\langle s_j^{l}|\rho_j|s_j^{l^{\prime}}\rangle $ do not depend on $j$.  
Hence  for each index $k$ also the $N$ eigenstates   $|  \psi_1^{k}\rangle, \ldots , |  \psi_N^{k}\rangle $, belonging each to a different density operator, represent a set of linearly independent symmetric states. 

 Let us introduce the $N$-dimensional subspace ${\cal H}_N^{k}$ spanned by the eigenstates  $|  \psi_1^{k}\rangle, \ldots,  |  \psi_N^{k}\rangle $  and let $I^{(k)}$ be the identity operator   in this subspace.  Since the relation    $\langle \psi_i^{k}|  \psi_j^{k^{\prime}}\rangle = 0$  ($k\neq k^{\prime}$) holds true for any $i,j$,  the $D$ subspaces  ${\cal H}_N^{k}$ with $k=1,\ldots,D$ are mutually orthogonal, which  implies that  $\sum_{k=1}^D I^{(k)} = I$. In the following we use the convention that operators labeled by the superscript $(k)$ act in the subspace ${\cal H}_N^{k}$. Introducing the symmetry operator  $V^{(k)}$ with   $ |\psi_{j}^{k}\rangle = V^{(k)} |\psi_{j-1}^{k}\rangle$, it follows that $\rho_j = V \rho_{j-1} V^{\dag},$ where   $V=V^{(1)}\otimes \ldots \otimes V^{(D)}$. Clearly, the $N$ mixed states given by Eq. (\ref{ND1}) or, equivalently, by  Eq. (\ref{ND3})  are symmetric, but in the following we do not necessarily assume that they occur with equal prior probabilities $\eta_j$.

The optimum measurement, discriminating the mixed states with a maximum probability of correct results for a fixed value of the failure probability  $Q$, is determined by the optimality conditions,    Eqs. (\ref{Q2}) and (\ref{Q4}), where because of Eq. (\ref{ND3}) $ \rho =  \sum_{k=1}^D \lambda_k\,  \rho^{(k)}$ with $\rho^{(k)}=\sum_{j=1}^N \eta_j\,\rho_j^{(k)}$.
Using the orthogonality of the different subspaces  ${\cal H}_N^{k}$ it follows that  the optimality conditions are satisfied  for a certain value of the real multiplier $a$ and for certain operators  $Z$ and  $\Pi_i$  ($i=0,1,\ldots,N$) when  
\begin{equation}
\label {ND5}
 Z =\sum_{k=1}^D \lambda_k\, Z^{(k)},\quad   \Pi_i=\sum_{k=1}^D \Pi_i^{(k)}\;\; {\rm with} \;\sum_{i=0}^N \Pi_i^{(k)}  = I^{(k)},
 \end{equation}
 provided that  $Z^{(k)}$ and  $\Pi_i^{(k)}$  ($i=0,1,\ldots,N$) satisfy the corresponding optimality conditions  in their respective subspaces  ${\cal H}_N^{k}$ for the same value of $a$.  The latter requirement can indeed be fulfilled in our problem, since  in each of the different subspaces ${\cal H}_N^{k}$  the structure of the resulting optimality conditions is identical,  due to the fact that  $\langle \psi_i^{k}|  \psi_j^{k}\rangle $ does not depend on $k$. In particular,  this means that in the optimum measurement ${\rm Tr\,}(\rho_j^{(k)}\Pi_j^{(k)} ) = {\rm Tr\,}(\rho_j^{(1)}\Pi_j^{(1)})$   for $k=2,\ldots, D$. Consequently, using again the orthogonality of the different subspaces ${\cal H}_N^{k}$,  we find that the detection operators given in Eq. (\ref{ND5}) 
maximize $P_c$ at the fixed failure probability  $ Q={\rm Tr\,}(\rho\Pi_0)$ that can be written as  $Q= \sum_{k}\lambda_k\, {\rm Tr\,}(\rho^{(k)}\Pi_0^{(k)} ) = {\rm Tr\,}(\rho^{(1)}\Pi_0^{(1)})$.   According to Eq. (\ref {Qdual5}) they yield the maximum probability of correct results 
\begin{equation}
\label {ND6}
P_c^{max}\big|_Q \!=\!\sum_{k=1}^D \!\lambda_k \sum_{j=1}^N \!\eta_j{\rm Tr}\left(\!\rho_j^{(k)}\Pi_j^{(k)}\!\right )
\!=\!\sum_{j=1}^N \eta_j{\rm Tr}\left(\rho_j^{(1)}\Pi_j^{(1)}\right ).
\end{equation}
The dependence of $P_c^{max}\big|_Q$ on $Q$ is  thus exactly the same as the respective dependence that arises from  the corresponding discrimination problem for the states $|\psi_1^{k}\rangle, \ldots, |\psi_N^{k}\rangle $ in any one of the orthogonal subspaces ${\cal H}_N^{k}$ $(k=1,\ldots,D)$. Without lack of generality in Eq. (\ref{ND6}) we referred to the subspace  ${\cal H}_N^{1}$. In other words,  provided that the pure-state optimization problem   can be solved for the set of linearly independent symmetric states $\{|\psi_j^{1}\rangle \} $ occurring with the prior probabilities $\eta_j$ $(j=1 \ldots,N)$, that is when the optimum detection operators in the subspace ${\cal H}_N^{1}$ can be determined, we know the complete solution. 

We  emphasize that $P_c^{max}\big|_Q$  does not depend on the rank $D$ of the mixed states $\rho_j$ nor on their matrix elements $s_{ll^{\prime}}$, or their eigenvalues $\lambda_k$, respectively.   
Using the pure-state results derived in this paper, we obtain analytical solutions in two cases:\\
\noindent (i) For  $N$ mixed states that are defined  by Eq. (\ref{ND1}) or, equivalently, by  (\ref{ND3}) and  occur with equal prior probabilities  $\eta_j=1/N$, the final result $P_c^{max}\big|_Q$ is  represented by Eq.  (\ref{sol-2})  with $d=N$, together with  Eqs. (\ref{sol-5a}) and  (\ref{sol-5b}). The failure probability  $Q^{\prime}$, given by  $Q^{\prime}=S$ if $S\geq 0$ and $Q^{\prime}= (N-1)|S|$  if $ - \frac{1}{N-1}\!< S \leq 0$, now corresponds to the minimum failure probability necessary for unambiguously  discriminating the $N$ mixed states.\\
\noindent (ii) When  $N=2$ and   $\eta_1 \leq \eta_2$ the final result $P_c^{max}\big|_Q$ for optimally discriminating the two mixed states described  by Eqs. (\ref{ND1}) or  (\ref{ND3}), respectively, is  obtained by substituting the expressions  $\rho_{11}=\eta_1$,  $\rho_{22}=\eta_2$,  $\rho_{12}=\sqrt{\eta_1\eta_2}\,S$  and $C=1$ into Eqs. (\ref{c6}) -- (\ref{c10}), thus using the corresponding  solution for the optimum discrimination of two pure states occurring with arbitrary prior probabilities.  We note that two pure states can be always written as symmetric states with respect to a suitable basis. 
Our general  result for $P_c^{max}\big|_Q$ reveals that the minimum failure probability $Q^{\prime}$ required for unambiguously discriminating the two mixed states is given by Eq. (\ref{limit6}), in agreement with earlier results for the optimum unambiguous discrimination of these two special mixed states \cite{BFH,herzog}.

\section{Further applications}

\subsection{$N$ equiprobable mixed qudit states resolving the identity operator}

When the dimension $d$ of the joint Hilbert space is larger than two and the states are genuinely mixed, that is when the discrimination problem cannot be reduced to the problem of discriminating pure states, it is in general hard to obtain analytical solutions. However, there is an exceptional case.  
We consider $N$ states that occur with equal prior probabilities, $\eta_j=1/N$,  and are described by the special density operators $\rho_{j}$ $(j=1,\ldots, N)$, where   
\begin{equation}
\label {O1}
 \rho_{j}= p\,|\psi_j\rangle\langle \psi_j|+
\frac{1-p}{d}I \quad {\rm with} \;\; \rho=  \frac{1}{N}\sum_{j=1}^N  \rho_j = \frac{I}{d}.
\end{equation}
Here $I$ is the identity operator  in  ${\cal H}_d$, and  $0\leq p \leq 1$.  Eq. (\ref{O1}) 
means that  the identity operator  can be resolved as a weighted sum over the density operators $\rho_j$.  
The largest eigenvalue of any one of the operators $ \tilde{\rho}_j=   \rho^{-1/2} \eta_j\rho_j   \rho^{-1/2}= \frac{d}{N} \rho_j$,
determining the maximum confidence $C_j$ of the outcome $j$, is given by $(pd+1-p)/N$, and the corresponding eigenstate is $|\psi_j\rangle$. Taking into account that $\rho$ is proportional to $I$, it follows from the considerations in Sc. II B that for maximum-confidence discrimination both the operators 
$\bar{\Pi}_j$ and  $\Pi_j=\rho^{-1/2}\bar{\Pi}_j \rho^{-1/2}$ are proportional to $|\psi_j\rangle\langle \psi_j|$. Since Eq. (\ref{O1})
implies that 
$\frac{1}{N}\sum_{j=1}^N  |\psi_j\rangle\langle \psi_j| = \frac{I}{d}$,
the operators 
\begin{equation}
 \label {O3}
 \Pi_0=0, \quad      \Pi_j= \frac{d}{N} |\psi_j\rangle\langle \psi_j| \quad (j=1,\ldots,N) 
\end{equation}
fulfill the completeness relation. Therefore maximum-confidence discrimination is possible without inconclusive results, and we get with the help of Eq. (\ref{conf1})
\begin{equation}
 \label {O2}
C_1 =\dots =C_N\equiv C= \frac{1+p(d-1)}{N}, \quad  Q^{\prime}=Q_{min}^{MC}=0.
\end{equation}
From Eq. (\ref{lim2}) we then obtain 
\begin{equation}
 \label {O2a}
P_c^{max}\big|_Q =  C(1-Q)\quad {\rm if}  \;\;Q \geq 0, 
\end{equation}
which in particular means that 
$P_c^{max}\big|_{Q=0}=\frac{1+p(d-1)}{N}$.  
The detection operators given by Eq. (\ref{O3}) describe 
optimized maximum-confidence discrimination since they yield the smallest possible failure probability, $Q=0$. On the other hand, they also describe  minimum-error discrimination, since they maximize the probability of correct results when $Q$ is fixed at the value  $Q=0$. Hence  both measurements coincide and the relative rate of correct results cannot be increased by admitting inconclusive results.
 
 Eqs. (\ref{O3}) -  (\ref{O2a}) agree with  our previous results  \cite{herzog2} obtained by studying  the optimized maximum-confidence discrimination of symmetric states obeying  Eq. (\ref{O1}). However, they  show that these results also hold in a more general case. While the derivation in  \cite{herzog2}  supposes  symmetric states $|\psi_j\rangle$ $ (j=1,\ldots, N)$ with equal  expansion coefficients $c_l$ with respect to the basis of the symmetry operator, $c_l=1/\sqrt{d}$ for $l=1,\ldots,d$ (see also Appendix B), we emphasize that in Eq. (\ref{O1}) the $N$ states $\rho_j$, or  $|\psi_j\rangle$, respectively, need not necessarily be symmetric.

Moreover, the expression for $P_c^{max}\big|_{Q=0}$  following from Eqs. (\ref{O2}) and (\ref{O2a}) generalizes the known result  for  minimum-error discrimination of pure states  resolving the identity operator  \cite{yuen,clarke} to a special class of mixed states.  We mention that the corresponding pure-state measurement  has been experimentally realized for a set of qubit states ($d=2$) with $N=3$, given by the symmetric trine states, and also for a set with  $N=4$, given by the tetrad states \cite{clarke}.   The latter  are defined as     
\begin{equation}
 \label {O3a}
|\psi_{j}\rangle \!=\!\frac{\sqrt{2}\;
{\rm exp}\!\left(\frac{ 2\pi i}{3}j\right)\!  |1\rangle - |0\rangle }{\sqrt{3}} \;\;(j=1,2,3),\;\;|\psi_4\rangle\! =\!|0\rangle.
\end{equation}
Note that the tetrad states do not belong to the class of fully symmetric states considered in the previous section, but possess only a partial symmetry. It is easy to check that they  fulfill the requirement $\frac{1}{4}\sum_{j=1}^4  |\psi_j\rangle\langle \psi_j| = \frac{I}{2}$. When $p<1$ the  end points of  the Bloch vectors belonging to the four depolarized tetrad states resulting from Eq.  (\ref{O1}) together with Eq. (\ref{O3a})  form a regular tetrahedron within the Bloch sphere, with circum radius $p$ and edge length $p\sqrt{8/3}$. In the special case $N=4$ and $d=2$ we get  $P_c^{max}\big|_{Q=0}= (1+p)/4$, in accordance with the  result following from a recent solution  \cite{jafarizadeh} for the  discrimination of mixed qubit states with Bloch vectors forming a regular polyhedron, where an approach to study minimum-error discrimination was applied that  is based on a geometrical method using Helstrom families of ensembles in convex optimization \cite{kimura,jafarizadeh}.

\subsection{Partially symmetric states}

The methods developed in this paper can be extended to the problem of discriminating between states that possess a certain partial symmetry. We assume that the given set of  states  consists of two sets of a equiprobable symmetric states and that  the total density operators resulting from each set commute.  More precisely, we refer to the discrimination of $N$ states $(N > M\geq 1)$, where the states $\rho_1,\ldots \rho_M$ are symmetric as described in Sec. IV, occurring with equal prior probabilities $\eta_1/M$, and where the remaining states $\rho_{M+1},\ldots \rho_N $ with equal prior probabilities  $(1-\eta_1)/(N-M)$ are also symmetric. In addition, we assume that the operators $\sum_{j=1}^M \rho_j$ and $\sum_{j=M+1}^N \rho_j$ have the same eigenbasis, which implies that the symmetry operators $V$ and $U$, referring to the two symmetric sets, are both diagonal in this eigenbasis. In analogy to the derivation of Eqs. (\ref{opt-symm1}) and (\ref{opt-symm2}) it follows that in the optimum measurement the detection operators can be supposed to obey the same symmetry as the density operators,  that is, $\Pi_{j}= V^{(j-1)}\Pi_1 V^{\dag( j-1) }$ for $j=1,\ldots,M$ and  $\Pi_{M+j}= U^{(j-1)}\Pi_{M+1} U^{\dag(j-1) }$ for $j=1,\ldots,N\!-\!M$.  
Eqs. (\ref{Q2}) and (\ref{Q4}) then reduce to the optimality conditions
\begin{eqnarray}
\label {part1}
&&Z-a\rho \geq 0,\;\;\quad\quad\quad (Z-a\rho)\;\Pi_0 = 0,\\
\label {part2}
 &&Z-\frac{\eta_1}{M}{\rho_1} \geq 0,\quad\quad\;  \left(\!Z-\frac{\eta_1}{M}{\rho_1}\right)\!\Pi_1=0, \\
\label {part3}
&& Z-\frac{\eta_2 \rho_{M+1} }{N\!-\!M}{} \geq 0,\quad  \left(\!Z-\frac{\eta_2\rho_{M+1}}{N\!-\!M}{}\right)\!\Pi_{M+1}=0,\;\;\;\qquad
\end{eqnarray}
where $\eta_1+\eta_2=1$. These conditions contain only  $\Pi_0$ and the two detection operators belonging  to the reference states $\rho_{1}$ and $\rho_{M+1}$ of the two symmetric sets.

\section{Relation to state discrimination with a  fixed  error probability}

\subsection{General considerations}

So far we considered the measurement strategy that maximizes the overall probability of getting a correct result, $P_c$, 
with a fixed value of the failure probability $Q$. The general relation 
 $P_c + P_e +Q =1$  implies that    
\begin{equation}
\label {PcmaxQ}
P^{max}_{c}\big|_Q =1-Q-P_e (Q) ,
\end{equation}
where  $P_e (Q)$ is the minimum overall error probability that can be obtained at the same fixed value of $Q$. 
 Another discrimination strategy maximizes $P_c$ under  the constraint that  the overall error probablity $P_e$ has a fixed value  \cite{hayashi-err,sugimoto}. 
We then get  
\begin{equation}
\label {PcmaxE}
\left.P^{max}_{c}\right|_{P_e}     = 1- P_e- Q(P_e),
\end{equation}
where $Q(P_e)$  is the minimum failure probability necessary to achieve the same fixed  error rate $P_e$. 
 
Let us investigate the relation between the optimization problems posed by these two strategies.  For this purpose we suppose that $P_e(Q)$, introduced in Eq. (\ref{PcmaxQ}),  is a monotoneously decreasing function of $Q$ in a certain interval around a value $Q=Q_{\alpha}$. Then  from  the assumption $P_e(Q_{\alpha})=P_e^{\alpha}$ (1) it follows that   $Q(P_e^{\alpha})=Q_{\alpha}$ (2),  where  $Q(P_e)$ is introduced in Eq. (\ref{PcmaxE}). 
In order to verify this intuitive statement,  we use an indirect proof.   Suppose that  $Q(P_e^{\alpha})=Q_{\beta}$ where  $Q_{\beta} > Q_{\alpha}$.  This means that  the failure probability $Q_{\alpha}$ is not large enough  to achieve the value $P_e^{\alpha}$, or, in other words, when $Q$ is fixed at $Q_{\alpha}$ we get a minimum overall error probability $P_e$ that is larger than $P_e^{\alpha}$, in contradiction to the assumption  (1). Now  suppose that  $Q(P_e^{\alpha})=Q_{\beta}$ where  $Q_{\beta} < Q_{\alpha}$. This means that the value $P_e^{\alpha}$ can be already reached at a value of $Q$ that is smaller than $Q_{\alpha}$  which together with the assumption (1)  is a contradiction to the fact that $P_e(Q)$ is monotonously decreasing. Hence the conclusion (2) indeed follows from the assumption (1). In an analogous way the equation   $P_e(Q_{\alpha})=P_e^{\alpha}$ can be shown to follow from $Q(P_e^{\alpha})=Q_{\alpha}$.  These findings are summarized as 
\begin{equation}
\label {arrow}
P_e(Q_{\alpha})=P_e^{\alpha}\;\;\leftrightarrow\; \;Q(P_e^{\alpha})=Q_{\alpha}.
\end{equation}
Taking Eqs. (\ref{PcmaxQ}) and (\ref{PcmaxE}) into account, we thus obtain the relation   $P^{max}_{c}\big|_{Q_{\alpha}}= 
 P^{max}_{c}\big|_{P_e^{\alpha}}$. Hence we conclude   that the detection operators maximizing $P_c$  at the fixed failure probability $Q_{\alpha}$ (or minimizing $P_e$ at this value $Q_{\alpha}$, respectively), are the same as the detection operators  maximizing $P_c$ at  the fixed error probability $P_e^{\alpha}=1- Q_{\alpha} - P^{max}_{c}\big|_{Q_{\alpha}}$ 
(or minimizing $Q$ at this value $P_e^{\alpha}$, respectively). 

The latter conclusion can be also obtained in more formal terms. 
 Let us assume that the detection operators $\Pi_0, \Pi_1,\ldots,\Pi_N$, the operator $Z$ and the scalar multiplier $a$  fulfill the optimality conditions, Eqs. (\ref{Q2}) and  (\ref{Q4}). Then the detection operators determine the measurement that  maximizes $P_c$ at the fixed failure probability $Q=a^{-1}  {\rm Tr\,}(Z\Pi_0)$, as becomes obvious from  Eq. (\ref{Qdual5a}). This yields  the probability of correct results $\left.P^{max}_{c}\right|_{Q}=1-P_e- Q= {\rm Tr\,}Z-aQ$, see Eq.  (\ref{Qdual5a}).  From the latter equality it follows that in the optimum measurement $P_e=1- {\rm Tr\,}Z- (1-a)Q$.  Taking again  Eq.  (\ref{Qdual5a}) into account, we conclude that the same optimum detection operators also characterize a measurement which maximizes $P_c$ at the fixed overall error probability $P_e= 1- {\rm Tr\,}Z- ( a^{-1}-1)  {\rm Tr\,}(Z\Pi_0)$. 

Due to the connection between the optimization problems in the two strategies we can directly determine  the solution $\left.P^{max}_{c}\right|_{P_e}$ from the solution  $\left.P^{max}_{c}\right|_{Q}$  and vice versa, using Eqs. (\ref{PcmaxQ}) and (\ref{PcmaxE})  and taking into account that  according to Eq. (\ref{arrow}) the function $Q(P_e)$ is the inverse of the function $P_e(Q)$.  
We note that the determination of  
  $\left.P^{max}_{c}\right|_{P_e}$  is only of practical interest  if  
\begin{equation}
\label {PE}
P_e \leq \left. P_E = 1-P^{max}_{c}\right|_{Q=0}  
\end{equation}
with  $P_E$ denoting the minimum error probability  obtainable in  the strategy of minimum-error discrimination \cite{helstrom,holevo}, where  inconclusive results do not occur. Admitting a larger value of $P_e$ does not yield any advantage since for  $P_e > P_E$ we get   $\left.P^{max}_{c}\right|_{P_e}\! \leq 1-P_e <1-P_E$, where  Eq. (\ref{PcmaxE}) has been used.    

\subsection{Example}

 In the following we present an example where we derive the maximum probability of correct results with a fixed error rate, $\left.P^{max}_{c}\right|_{P_e}$, with the help of  the  result for $\left. P^{max}_{c}\right|_{Q}$.  We consider the discrimination of  two mixed qubit states with $C_1=C_2=C$, see Sec. III C, assuming that $\rho_{11}\leq \rho_{22}$. 
From Eq. (\ref {PcmaxQ}) together with Eqs.  (\ref{c6}) and   (\ref{c8})  we find that  
\begin{equation}
\label {PcmaxE1}
P_e (Q)=  \!\frac{1-Q}{2}-\frac{2C-1}{2}\sqrt{ ({1}-2|\rho_{12}|) ({1}+2|\rho_{12}|-2Q)},
\end{equation}
provided that $Q$ is restricted to a certain interval. Inserting the boundaries of this interval into Eq. (\ref{PcmaxE1}) leads to  the restriction $P_e^{\prime} \leq P_e(Q) \leq P_E$, where  $P_E$ refers to minimum-error discrimination, see Eq. (\ref{PE}), and  
\begin{equation}
\label {equiv1a}
P_e^{\prime} =\left \{
\begin{array}{ll} \!   P_e(2 |\rho_{12}|)= (1-C)(1-2 |\rho_{12}|) \; & \mbox{if $\;
            |\rho_{12}| \leq \rho_{11},   $}  \quad\\
\! P_e (Q_ {cr}) \; \; & \mbox{if $\;
               |\rho_{12}| \geq  \rho_{11} $}
 \end{array}
\right.
\end{equation}
with \mbox{$P_e(Q_ {cr})\!=\!  \frac{(\rho_{11}- |\rho_{12}|)^2}{1-2|\rho_{12}|}+(1-C)(\rho_{22} -\rho_{11})$}.
From the function $P_e (Q)$ we can derive  the inverse function  $Q (P_e)$.  
Making use of  Eq. (\ref {PcmaxE}) we obtain  
\begin{eqnarray}
\label {equiv0}
&&\!\!\!\! \left. P_c^{max}\right|_{P_e}=
P_e  + (2C \!-\!1)^2(1-2|\rho_{12}|)\qquad\qquad\qquad\ \\
&&\!\!\!\!\quad+2(2C\!-\!1) \sqrt {(1-2|\rho_{12}|)[P_e\!-\! C(1\!-\!C) (1-2|\rho_{12}|)]},\nonumber\\
&&\!\!\!{\rm if}\;\;   P_e^{\prime} \leq P_e \leq P_E, \qquad\qquad\qquad\quad \nonumber
\end{eqnarray}
which  reduces to $ \left.P_c^{max}\right|_{P_e}= ( \sqrt{P_e}\!  +\! \sqrt{1-2|\rho_{12}|})^2$ when $C=1$, that is when the states are pure, in accordance with \cite{sugimoto}.
Next we use  Eq.  (\ref {PcmaxQ})  together with Eq. (\ref{limit5}) and get 
$P_e(Q)= (1-C)(1-Q)$ for  $\;Q^{\prime}\leq Q\leq 1$,
where $Q^{\prime}$ is given by Eq. (\ref{limit6}). This restricts $ P_e(Q)$ to the interval  $0 \leq  P_e(Q) \leq P_e^{\prime\prime}$ with  
\begin{eqnarray}
\label {equiv8}
P_e^{\prime\prime} =\left \{
\begin{array}{ll} \! (1-C)(1-2 |\rho_{12}|)=P_e^{\prime}\; & \mbox{if $\;
            |\rho_{12}| \leq \rho_{11},    $}  \quad\\
\! (1-C)(1-Q_1)\; \; & \mbox{if $\;
               |\rho_{12}| \geq \rho_{11}. $}\quad\\
 \end{array}
\right.
\end{eqnarray}
For $C\neq 1$  we can determine the inverse function $Q(P_e)$ and insert it  into  Eq.  (\ref {PcmaxE}),  arriving at
\begin{equation}
\label {equiv8}
\left. P_c^{max}\right|_{P_e} =   P_e \frac{C}{1-C}  \quad \mbox{if  $\;0\leq P_e\leq P_e^{\prime\prime}$\quad($C\neq 1$). }
\end{equation}
Equation (\ref{equiv8})  reflects the fact that unambiguous discrimination, where $P_e=0$, is impossible when $C\neq 1$, since then   $P_c^{max}=0$ which means that $Q=1$ and the measurement  always fails.  

If  $  |\rho_{12}| \leq \rho_{11}$  Eqs. (\ref{equiv0}) --   (\ref{equiv8})  determine  the complete solution since in this case    $P_e^{\prime}= P_e^{\prime\prime}$. For $C=1$ the solution is in agreement with the  result obtained for pure states by directly performing the optimization when $P_e$ is fixed
   \cite{sugimoto}.  

If $  |\rho_{12}| \geq \rho_{11}$, that is if   $P_e^{\prime}\neq P_e^{\prime\prime}$, we still have to consider the  interval $ P_e^{\prime\prime}= P_e(Q_1)\leq P_e\leq  P_e(Q_ {cr})=P_e^{\prime}$. 
From  Eqs.   (\ref{PcmaxQ}) and  (\ref {s33}) we find that 
\begin{equation}
\label {equiv3}
P_e(Q)= (1- C)(1-Q) + (2C-1)\gamma_1(Q) 
\end{equation}
if $Q_ {cr} \leq Q \leq Q_1$, where  $\gamma_1(Q_1)=0$.   
Using Eq. (\ref{s32}) it is in principle possible to  invert Eq. (\ref{equiv3}), that is, to determine  the function $Q(P_e)$, which yields  $P_c^{ max}|_{P_e}$ according to   Eq.  (\ref {PcmaxE}). Since for $C\neq1$  the calculations are rather involved, we specialize to the case $C=1$, where $P_e^{\prime\prime}=0$  and $P_e(Q)=\gamma_1(Q)$,
which due to  Eq. (\ref {PcmaxE}) yields   $P_c^{max}|_{P_e}=1- \gamma_1-Q(\gamma_1)$.   The right-hand side of the latter equation is equivalent to $ \gamma_2$, as becomes obvious from Eq. (\ref{s32}).  Substituting $\gamma_1=P_e$ into the equation 
 for $ \gamma_2$ given by Eq. (\ref{s32}), we obtain  for $  |\rho_{12}| \geq \rho_{11}$   
\begin{eqnarray}
\label {equiv5}
&&\!\!\!\!\! \left.P_c^{max}\right|_{P_e}\! =\! \frac{1}{\rho_{11}^2}     \left(|\rho_{12}|\sqrt{P_e}\!+\! 
\sqrt{\Delta} \sqrt{\rho_{11}-P_e}\right)^2\;\quad\\
&&\!\!\!\!\!\! \mbox{ if $\;C=1\;$ and  $\;0\leq P_e\leq   \frac{(\rho_{11}- |\rho_{12}|)^2}{1-2|\rho_{12}|}=    P_e^{\prime}$},                   \nonumber 
\end{eqnarray}
where $\Delta=\rho_{11}\rho_{22}\! -\! |\rho_{12}|^2$. 
Equation (\ref {equiv5}) together with  Eq. (\ref{equiv0}) for $C=1$  determines the complete solution when the states are pure  with $  |\rho_{12}| \geq \rho_{11}$. Using Eq. (\ref{s11}) we find that this solution is  in agreement with the results obtained  for two pure states by directly performing the  optimization for a fixed value of $P_e$ \cite{sugimoto}. 

\section{Summary and Conclusions}

In the main  part of the paper we considered a measurement for state discrimination  that  minimizes the error probability $P_e$, or maximizes the probability $P_c$ of correct results,  respectively, when a certain fixed probability  $Q$ of inconclusive results  is admitted. For a number of problems not treated before we derived analytical solutions for the optimum measurement: \\
 \noindent 
 (i) We investigated  the discrimination of two arbitrary mixed qubit states that occur with arbitrary prior probabilities.  For the case that the two conclusive outcomes can be discriminated with the same maximum confidence we obtained the complete solution, see Eqs. (\ref{c6}) - (\ref{c10}). This solution includes the discrimination of two pure states occurring with arbitrary prior probabilities.   \\
 \noindent 
(ii) We studied the discrimination of $N$ symmetric states spanning a $d$-dimensional Hilbert space. 
For a certain class of symmetric equiprobable pure qudit states $(d\leq N)$ we derived the  solution, given by Eqs. (\ref{sol-2}) and  (\ref{sol-3}). As a special case, this solution  contains  the  discrimination of $N$ symmetric linearly independent pure states with equal mutual overlaps, see Eqs. (\ref{sol-5a}) and  (\ref{sol-5b}). Moreover, we also obtained  the solution for the discrimination of $N$ symmetric equiprobable mixed qubit states, given by Eqs. (\ref{sol-7}) and  (\ref{sol-8}), and of $N$ special  symmetric mixed states of rank $D$ spanning a joint Hilbert space of dimension $ND$, see Eq. (\ref{ND6}). \\
 \noindent 
 (iii) We solved the optimization problem for a  case of mixed qudit states that are complete in the sense that a weighted sum of their density operators is equal to the identity operator, and we also specified the general optimality conditions for a certain  kind of partially symmetric states.  The treatment of the optimization problem resulting for the latter case is left for further investigations. 

In the final part of the paper we showed that there exists a general relation between the solutions for optimum state discrimination in the two different  discrimination strategies where either the rate $Q$ of inconclusive results,  or  the overall error rate $P_e$, has a fixed value. This relation, expressed by Eqs.  (\ref{PcmaxQ}) - (\ref{arrow}),  holds for an arbitrary  number $N$ of mixed states. It implies that by solving the optimization problem in one of the two strategies, one can also obtain the solution in the other strategy. As an illustration we presented an example where for two mixed qubit states   the maximum rate of correct results with a fixed error rate  $P_e$  is derived from the solution for optimum state discrimination with a fixed rate $Q$ of inconclusive results. 

In order to solve the optimization problem for a fixed probability  $Q$ of inconclusive results, we applied the  operator conditions \cite {fiurasek} determining the optimum measurement. As discussed in our paper, these optimality conditions,
  Eqs. (\ref{Q2}) and  (\ref{Q4}),  provide a very general approach  for treating  various  optimized state discrimination measurements, as far as only overall  probabilities, averaged over all outcomes,  are considered. In the appropriate limiting cases, they  describe  minimum-error discrimination,  on the one hand, while on the other hand   they refer to optimized maximum-confidence discrimination provided that the maximum confidence is the same for each conclusive outcome,      or to optimum unambiguous discrimination, respectively,  in the special case when the maximum confidence is equal to unity. 

We note  that related work has been done independently by E. Bagan and R. $\rm Mu\tilde{n}oz$-Tapia (Barcelona) and  G. A. Olivares-Renter\'{i}a  and J. A. Bergou  (New York) \cite{bagan}.

\begin{acknowledgments}
Discussions on various aspects of state discrimination with J\'{a}nos Bergou (Hunter College, New York)  and Oliver Benson  (Humboldt-Universit\"at, Berlin) are gratefully acknowledged.  

\end{acknowledgments}

\appendix

\section{}

Here we present the derivation of the optimum measurement  that discriminates two mixed qubit states 
when $C_1=C_2=C$.  Eq. (\ref{s13a}), valid for $\Gamma - aI > 0$, yields    
\begin{equation}
\label {s21}
 \Gamma_{22}=\Gamma_{11}=\frac{a^2-C(1-C)}{2a-1},\quad  |\Gamma_{12}|=\Gamma_{11}-a,
\end{equation}
where  we have to require that $\Gamma_{11}>a$ and 
$\Gamma_{11}>C>1/2$ to ensure the positivity of the expressions under the square-root signs in Eqs. (\ref{s14a}) -- (\ref{s14c}).   
From Eqs. (\ref{s18a}) and (\ref{s18c}) we obtain  the relations 
\begin{equation}
\label {s22}
e^{i\delta}= - e^{i\phi}, \quad \beta_1+\beta_2=\frac{1-Q}{2\Gamma_{11}-1}=\frac{2|\rho_{12}|-Q}{|\Gamma_{12}|}.
\end{equation}
Using Eq. (\ref{s21}) in the last equality of  Eq. (\ref{s22}) we get  
\begin{equation}
\label {s23a}
a= \frac{1}{2}+ \frac{2C-1}{2}\sqrt{\frac{1-2|\rho_{12}|}{1+2|\rho_{12}|-2Q}}.
\end{equation}
 After calculating  the matrix elements $\Gamma_{11}$ and $\Gamma_{22}$ that result for this value of $a,$
 we obtain from  Eqs. (\ref{s18a}) and    (\ref{s18b})  the solutions    
\begin{equation}
\label {s24}
\beta_{1/2}= \frac{\sqrt{(1-2|\rho_{12}|)(1+2|\rho_{12}|-2Q)} \pm(\rho_{11}-\rho_{22})}{2(2C-1)}.
\end{equation}
 Due to  Eqs. (\ref{s22}) and  (\ref{s24})   a solution with positive constants   $\beta_{1}$ and $\beta_{2}$ can only result when  the two conditions
\begin{eqnarray}
\label {s24a}
Q \leq  2|\rho_{12}|, \quad
 Q \leq \frac{\rho_{11}  \rho_{22}- |\rho_{12}|^2} {\frac{1}{2}-|\rho_{12}|} \equiv Q_ {cr} 
\end{eqnarray}
are simultaneously fulfilled. In  the second condition we used the relation   $(\rho_{11}-\rho_{22})^2=1-4\rho_{11} \rho_{22}$ and introduced the critical value $Q_{cr}$ of the failure probability. 
For later purposes  we note that 
\begin{eqnarray}
\label{s26} 
2|\rho_{12}| \leq Q_ {cr} \quad {\rm if}&&\!\!  |\rho_{12}| \leq {\rm min}\{\rho_{11}, \rho_{22}\},    \;\; 
\qquad\\ 
\label{s26a}
2|\rho_{12}| \geq Q_ {cr} \quad {\rm if}&& \!\! |\rho_{12}|\geq {\rm min}\{\rho_{11}, \rho_{22}\}.   \;\; 
\qquad
\end{eqnarray}
To see this,  let us assume  that  $ |\rho_{12}| \leq \rho_{11}\leq \frac{1}{2}$, which  leads to  
$\left(\frac{1}{2}-|\rho_{12}|\right)^2 \geq \left(\frac{1}{2}-\rho_{11}\right)^2 =\frac{1}{4}-\rho_{11}\rho_{22.}$
Hence it follows that  $ \rho_{11}  \rho_{22}- |\rho_{12}|^2 \geq   |\rho_{12}|-2 |\rho_{12}|^2 $ and consequently $2|\rho_{12}| \leq Q_ {cr}$. Analogous considerations hold for the other possible cases.   

When the two conditions in Eq. (\ref{s24a}) are met, we obtain from  Eq. (\ref{s11b}) the optimum detection operators 
\begin{equation}
\label {s24c}
\Pi_{i}= \beta_{i}   \rho^{-1/2} |\tilde{\pi}_i\rangle\langle\tilde{\pi}_i| \rho^{-1/2}\quad (i=1,2), 
\end{equation}
where the states $|\tilde{\pi}_i\rangle$ result from  Eqs. (\ref{s14b}) and  (\ref{s14c}),  
\begin{eqnarray}
\label {s24d}
|\tilde{\pi}_1\rangle \!=\sqrt{\frac{2C-1}{2}}\left( \sqrt{b +1 } |\nu_1\rangle + e^{-i\phi} \sqrt{b-1 } |\nu_2\rangle\right)\! \quad\\
\label {s24e}
|\tilde{\pi}_2\rangle \!=\sqrt{\frac{2C-1}{2}} \left( \sqrt{b -1 } |\nu_1\rangle + e^{-i\phi} \sqrt{b+1 } |\nu_2\rangle \right)\! \quad\\
\label {s24f}
{\rm with}\; \;b= \frac{1-Q}{ \sqrt{(1-2|\rho_{12}|)(1+2|\rho_{12}|-2Q)}}.\;\;\quad\,\qquad
\end{eqnarray}
For completeness, we also give an explicit expression for the failure operator.  
Because of  Eqs. (\ref{s11a}) and (\ref{s14a}) the latter can be written as 
\begin{equation}
\label {s25b}
\Pi_0=Q\rho^{-1/2} |\bar{\pi}_0\rangle \langle \bar{\pi}_0 | \rho^{-1/2}=\frac{Q}{Q_ {cr}} |{\pi}_0\rangle \langle {\pi}_0|, 
\end{equation}
 where  $|\pi_0\rangle =\sqrt{Q_ {cr}}{ \rho^{-1/2}{|\bar{\pi}}_0\rangle}$ with $\langle\pi_0|\pi_0\rangle=1.$ 
Here we took into account that 
\begin{equation}
\label {s25c}
|\bar{{\pi}}_0\rangle =\frac{ |\nu_1\rangle + e^{-i\phi} |\nu_2\rangle}{\sqrt{2}} \quad  {\rm and}\;\; 
 \langle \bar{\pi}_0 |\rho^{-1} | \bar{\pi}_0\rangle = 
\frac{1}{ Q_ {cr}},   
\end{equation}
where in the second equality use has been made of the matrix elements of $\rho^{-1}$ with respect to the basis $\{|\nu_1\rangle, |\nu_2\rangle\}$.  

If   $|\rho_{12}| \leq {\rm min}\{\rho_{11}, \rho_{22}\}$ and therefore $2|\rho_{12}|\leq Q_ {cr}$,  the condition  $Q\leq 2|\rho_{12}|$ is sufficient to guarantee that both inequalities in Eq. (\ref{s24a}) hold, which means that the detection  operators given  by  Eqs.  (\ref{s24c}) --  (\ref{s25b})  describe the  optimum measurement.  Using  Eq. (\ref{Qdual5}) together with the expressions  for $\tilde{\rho}_1$ and $\tilde{\rho}_2$ that ensue from Eqs. (\ref{s1a}) and  (\ref{s1b}) when $C_1=C_2=C,$ we obtain with the help of Eq. (\ref{s24c}) the maximum probability of correct results $P_c ^{(0)}$ given in Eq. (\ref{c7}).  After combining this result with  Eq. (\ref{limit5}) we arrive at  Eq. (\ref{c6}). 

 We now focus on the case  that $|\rho_{12}| \geq {\rm min}\{\rho_{11}, \rho_{22}\},$
where due to Eqs. (\ref{s24a}) and  (\ref{s26a})   the detection  operators given by  Eqs.  (\ref{s24c}) -- (\ref{s25b}) describe the optimum measurement only as long as  $Q\leq Q_{cr}$.  In order to derive  the solution for $Q\geq Q_ {cr}$, we assume without lack of generality  that $|\rho_{12}|\geq  \rho_{11},$ which implies that $\rho_{11} < \rho_{22}$ and that $P_c^{max}=  C(1-Q)$ if $Q\geq Q_1$, see Eqs. (\ref{abb}) and (\ref{limit5}).   Equation (\ref{s24}) reveals that in this case  the parameter $\beta_1$ vanishes for $Q=Q_ {cr}$, which means that  $ \Pi_1=0$.  Therefore if $Q_ {cr} \leq Q \leq Q_1$ we have to  search  for a projective measurement with 
\begin{equation}
\label {s28}
  \Pi_1=0,\quad \Pi_2= |\pi_2\rangle\langle \pi_2|, \quad  \Pi_0=I- \Pi_2 
\end{equation}
that maximizes $P_c$ at the fixed value $Q$ and yields  $P_c^{max}(Q_1)=  C(1-Q_1)$. Instead of using again the optimality conditions, we can solve the optimization problem directly,  due to the simple structure of the detection operators. For this purpose we make the convenient general ansatz 
\begin{equation}
\label {s28a}
|\pi_2\rangle = \sqrt{\gamma_1}\,\rho^{-1/2}|\nu_1\rangle + e^{i\chi}\sqrt{\gamma_2}\,\rho^{-1/2}|\nu
_2\rangle,
\end{equation}
 where the phase $\chi$ and the nonnegative parameters $\gamma_1$ and $\gamma_2$  have to be determined.  The normalization condition  $\langle\pi_2|\pi_2\rangle=1$ leads to  the constraint 
\begin{equation}
\label {s29}
\gamma_1 \rho_{22}\! +\! \gamma_2 \rho_{11}\! -\! 2  \sqrt{\gamma_1 \gamma_2}|\rho_{12}|\cos(\chi \!+\!\phi)\! =\!  \rho_{11}\rho_{22}\! -\! |\rho_{12}|^2\!\equiv \!\Delta,
\end{equation}
where we used the matrix elements of $\rho^{-1}$ with respect to the basis $\{|\nu_1\rangle, |\nu_2\rangle\}$ and took into account that  $ \rho_{12}=  |\rho_{12}| e^{i\phi} $. 
The fixed probability $Q$ of inconclusive results can be expressed as   
\begin{equation}
\label {s30}
Q= 1- {\rm Tr\,}(\rho\Pi_2)= 1-( \gamma_1 + \gamma_2).
\end{equation}
From $ P_c=  \eta_2 {\rm Tr\,}(\rho_2\Pi_2)= {\rm Tr\,}(\tilde{\rho}_2\rho^{1/2} \Pi_2\rho^{1/2})$ we obtain the probability of correct results, 
\begin{equation}
\label {s31}
P_c=(1-C)\gamma_1+ C \gamma_2,
\end{equation}
where   Eq. (\ref{s1b}) has been used. In order to maximize $P_c$  we need to allow for the largest possible values of $\gamma_1$ and $\gamma_2$ consistent with Eq. (\ref{s29}), that is we have to put $\cos(\chi \!+\!\phi)=1$,  which means that $\chi = -\phi$.  Using   Eqs. (\ref{s29}) and   (\ref{s30}) we then find that    
\begin{equation}
\label {s32}
{\gamma_2} = \frac{\left(|\rho_{12}|\sqrt{\gamma_1}+ 
\sqrt{\Delta} \sqrt{\rho_{11}-\gamma_1}\right)^2
}{\rho_{11}^2}=1-Q-\gamma_1.
\end{equation}
The equation arising from the second equality sign in Eq. (\ref{s32}) can be solved to yield an explicit expression  for $\gamma_1(Q)$.   
Making use of Eq.  (\ref{s31}) together with Eq. (\ref{s30}) we get the solution    
\begin{equation}
\label {s33}
P_c^{max}= P_c^{(1)}= C(1-Q)-(2C-1)\gamma_1(Q), 
\end{equation}
 which is explicitly given in Eq. (\ref{c10}).  Clearly, our derivation requires that  $\gamma_1(Q)\geq 0$. 
According to  Eq.  (\ref{s32}) the boundary case  $\gamma_1= 0$ implies that  $\Delta/\rho_{11}=1-Q$ and therefore  $Q=Q_1$,   
see Eq. (\ref{abb}). Hence Eq. (\ref{s33}) is valid for $Q_ {cr} \leq Q \leq Q_1$,  yielding  $P_c^{(1)}(Q_1)= C(1-Q_1)$, as required. Taking Eq. (\ref{limit5}) into account we arrive at the final result, Eq. (\ref{c8}).  A direct calculation shows that indeed   $P_c^{(0)}(Q_ {cr})= P_c^{(1)}(Q_ {cr})$. 

We still  consider the question under which condition the relation $C_1=C_2=C$ holds true when the mixed qubit  states are represented with the help of the general ansatz  $\rho_{1/2}\!=\! p_{1/2}|\psi_{1/2}\rangle\langle \psi_{1/2}|+\!\frac{1\!-\!p_{1/2}}{2}\,I$, see Eq. (\ref{c1}).   
After calculating  the spectral representations of $\tilde{\rho}_j=\rho^{-1/2}  \eta_j\rho_j   \rho^{-1/2}$ $(j=1,2)$ and comparing the result with  Eqs. (\ref{s1a}) and  (\ref{s1b})  we 
arrive at  the condition 
$(\eta_1 p_1)^2-(\eta_2 p_2)^2 = \eta_1-\eta_2$, leading to the explicit expression
\begin{equation}
\label {c5}
(2C\!-\!1)^2\!=\!1-\! \frac{2\sqrt{(1-p_1^2)(1-p_2^2)}}{1+ \sqrt{(1-p_1^2)(1-p_2^2)}+p_1 p_2(1-2|S|^2)}
\end{equation}
with  $S= \langle \psi_1|\psi_2 \rangle$.
Taking into account that   $ {\rm Tr\,}(\rho_1 \rho_2   )={\rm Tr\,}(\rho \tilde{\rho}_1  \rho \tilde{\rho}_2   )/( \eta_1\eta_2) $ we find with the help of   Eqs. (\ref{c1}), (\ref{s1a}) and   (\ref{s1b}) that   
\begin{equation}
\label {c5a}
\frac{1\!-\!p_1p_2(1\!-\!2|S|^2)}{2}
= \frac{C(1\!-\!C) \left ( \rho_{11}^2 \!+\!  \rho_{22}^2 \!-\!2 |\rho_{12}|^2 \right) \!+\!  |\rho_{12}|^2}{\eta_1(1-\eta_1)}.
\end{equation}
Together with  Eq. (\ref{eta1}) and with  the relation $\left(\eta_1^{-1}-1\right)^2=\frac{1-p_1^2}{1-p_2^2}$, following from the condition 
 for $C_1=C_2=C$,  Eqs. (\ref{c5}) and (\ref{c5a}) form a system of four equations for determining the  values of $\eta_1, p_1,  p_2$ and  $|S|$ when $C$ and the matrix elements $\rho_{ij}$ are given. 

\section{}

We consider the discrimination of  $N$ symmetric qudit states 
$\rho_{j}$  $(j=1,\ldots,N)$ described by  
\begin{equation}
\label {r10} 
\rho_{j}= p\,|\psi_j\rangle\langle \psi_j|+
\frac{1-p}{d}\,I \quad {\rm with}\quad |\psi_{j}\rangle = V^{(j-1)}|\psi_1\rangle, 
\end{equation}
where $0\leq p \leq 1$ and 
 $|\psi_1 \rangle= c_1 \!\sum_{l=1}^m  |r_l\rangle +  c_2 \sum_{l=m+1}^d |r_l \rangle,$
see Eq.  (\ref{sol-1}).    
Here $I$ is the identity operator in  the $d$-dimensional Hilbert space ${\cal H}_d$  spanned by  the $N$ pure states $|\psi_j\rangle$.  
Using the properties  of  the symmetry operator $V,$ given after Eq.   (\ref{t2}), 
  we can  determine  the spectral representation of  $\rho=\frac{1}{N}\sum_{j=1}^N {\rho_j}$, arriving at
\begin{equation}
\label {r11} \rho=r_{1}\sum_{l=1}^m
|r_l\rangle\langle r_l|+r_2\!\!\!\sum_{l=m+1}^d\!\!\!
|r_l\rangle\langle r_l|, 
 \quad
 r_{i} = p\,|c_{i}|^2+\frac{1-p}{d}
\end{equation}
with $i=1,2$. The structure of  $\rho_1$ suggests the ansatz 
\begin{equation}
\label {y0}
\Pi_1= |\tilde{v}\rangle\langle  \tilde{v}|\;\;\;\;{\rm with}\;\; |\tilde{v}\rangle = b_1 \sum_{l=1}^m  |r_l\rangle +  b_2\!\!  \sum_{l=m+1}^d \!\!\! |r_l\rangle. 
\end{equation}
 To be specific,  we  assume that  $0<|b_1|\leq  |b_2|.$ 
  We shall use the optimality conditions, Eqs. (\ref{opt-symm1}) and  (\ref{opt-symm2}), in order to investigate whether  this ansatz yields the optimum solution for the case that   $Z-a\rho > 0$.

 The equality in  Eq. (\ref{opt-symm1})  implies that $\Pi_0$ cannot span the whole $d$-dimensional Hilbert space if  $Z-a\rho > 0$. Using  Eqs.  (\ref{opt-symm3}) and (\ref{y0}) this  leads to  the requirement  
\begin{equation}
\label {opt-symm4}
N|b_2|^2=1,\quad  \Pi_0=\left(1-N|b_1|^2\right)\sum_{l=1}^{m}|r_l\rangle\langle r_l|,
\end{equation}
which arises  from  the positivity constraint $\Pi_0 \geq 0$ with  $|b_2| \geq  |b_1|$.  The resulting failure probability is given by
\begin{equation}
\label {opt-symm6}
Q={\rm Tr\,}(\rho\Pi_0) = (1-N|b_1|^2)m r_1.
\end{equation}
Taking into account that  $[Z,\rho]=0$ it becomes obvious from Eqs. (\ref{r11})  and (\ref{opt-symm4}) that the equality in  the  first  optimality condition, Eq. (\ref{opt-symm1}), holds true  if 
\begin{equation}
\label {opt-symm5}
Z= z_1\sum_{l=1}^m\! |r_l\rangle\langle r_l|+z_2\!\sum_{l=m+1}^d \!\!\!|r_l\rangle\langle r_l|\quad {\rm and}\;\;a= \frac{z_1}{ r_1}.
\end{equation}

Now we turn to the  equality in  the second optimality condition, Eq. (\ref{opt-symm2}), leading to   
$(NZ- {\rho}_1) |\tilde{v}\rangle \langle  \tilde{v}|=0$, which implies that   
$N\langle r_l|Z|\tilde{v}\rangle=\langle r_l|\rho_1|\tilde{v}\rangle$ for $l=1,\ldots,d$.  After  expressing  $\rho_1$  with the help of  Eqs. (\ref{r10}) and  (\ref{sol-1})  we arrive at  the  condition
\begin{equation}
\label {opt-symm7}
z_i= \frac{p}{N}\,\frac{c_i}{b_i}\left [ m c_1^{\ast}b_1
 +(d-m) c_2^{\ast}b_2\right]
+\frac{1-p}{dN} \quad(i=1,2).
\end{equation}
Since  $Z$ is Hermitian we choose the phase of $b_i$ to coincide with the phase of $c_i$ $(i=1,2)$.   Equations (\ref{opt-symm4}) and (\ref{opt-symm6}) then lead to   
\begin{equation}
\label {opt-symm8} 
 b_2\!=\!\frac{1}{\sqrt{N}}\frac{c_2}{|c_2|},\quad b_1\!=\!\frac{w}{\sqrt{N}}\frac{c_1}{|c_1|}\quad{\rm with}\;\; w=\sqrt{\frac{m r_1-Q}{m r_1}}.
\end{equation}
From Eq. (\ref{opt-symm7}) we obtain the relation     
\begin{equation}
\label {opt-symm8a} 
z_1\!-\!  \frac{1\!-\!p}{dN}\! =\!\frac{p|c_1c_2|}{N}\! \left (\!m \frac{|c_1|}{|c_2|}\!+\! \frac{d\!-\!m}{w}\!\right)= \frac{|c_1|}{w|c_2|}\left(\! z_2\!-\!  \frac{1\!-\!p}{dN}\!  \right),
\end{equation}
yielding explicit expressions for  $z_1$ and $z_2$.  Using   Eq. (\ref{opt-symm3a}) we get  the maximum  probability of correct results,  
$ P_c^{max}|_Q= \langle  \tilde{v}|\rho_1| \tilde{v}\rangle \equiv P_c^{(0)}$,  where because of    Eq. (\ref{opt-symm8})
\begin{equation}
\label {opt-symm12}  
P_c^{(0)}=\frac{p}{N}\big[m |c_1|w
+ (d-m)|c_2|\big]^2
+\frac{1\!-\!p}{Nd}(m w^2+ d\!-\!m).
\end{equation}
In order to decide whether this is indeed the optimum solution  we have to check whether the positivity constraints in Eqs. (\ref {opt-symm1}) and (\ref {opt-symm2}) are fulfilled.  

We start with the  
positivity constraint  in Eq. (\ref {opt-symm2}), which takes the form  
$Z-\frac{p}{N}|\psi_1\rangle\langle\psi_1|-\frac{1-p}{Nd}I \geq 0.$
Clearly, the constraint is satisfied provided that 
$Nd\langle\psi_1|Z|\psi_1\rangle \geq 1- p+pd$.
 After inserting the explicit representation of the operator $Z$, this inequality can be transformed into
\begin{equation}
\label {opt-symm18} 
m^2|c_1|^4+m(d\!-\!m)|c_1c_2|^2 \!\left(\!\frac{|c_1|}{w|c_2|}\!+\!\frac{w|c_2|}{|c_1|}\!\right)\!
+(d\!-\!m)^2|c_2|^4\!\geq \! 1.
\end{equation}
Due to the normalization condition in Eq.  (\ref{sol-1}) 
and to the fact that  $x+x^{-1}\geq 2$ for any $x$ it is obvious that Eq. (\ref{opt-symm18}) always holds true, which implies that the  positivity constraint in Eq. (\ref {opt-symm2}) is satisfied. 

The positivity constraint given in  Eq.  (\ref {opt-symm1}), on the other hand, which results from the fact that $Q$ is fixed and different from zero, imposes the relevant restriction.  Because of  Eq. (\ref{opt-symm5}) this constraint takes the form 
\begin{equation}
\label {opt-symm19} 
z_2-ar_2 = z_2- z_1\frac{r_2}{r_1}\geq 0,
\end{equation}
where $z_1$ and $z_2$ follow from  Eq. (\ref{opt-symm8a}).  It turns out that the constraint is satisfied in the following general  cases:\\ 
(i) When $p=1$ and therefore $r_i=|c_i|^2$ $(i=1,2)$,   Eq. (\ref{opt-symm19}) is fulfilled if   $w\geq |c_2|/|c_1|$, which with the help of   Eqs. (\ref{opt-symm8}) and    (\ref{sol-1}) leads   to the condition     
\begin{equation}
\label {opt-symm20}  
Q \leq  m(|c_1|^2-|c_2|^2)=1-d |c_2|^2 =Q^{\prime}.
\end{equation}
(ii) When $d=2$, and consequently  $m=d-m=1$, a straightforward calculation shows that  Eq. (\ref{opt-symm19})
holds true provided that $w^2\geq r_2/r_1$. Because of Eq.  (\ref{opt-symm8})  this yields   the requirement 
$Q \leq   r_1-r_2,$ or, using Eq. (\ref{r11}), 
\begin{equation}
\label {opt-symm21}  
Q \leq p(|c_1|^2-|c_2|^2)=p(1-2 |c_2|^2) = Q^{\prime}.
\end{equation}
 (iii) When $|c_1|^2=|c_2|^2= \frac{1}{d}$ for arbitrary values of  $p$ and $d$, implying that also 
$r_1=r_2= \frac{1}{d}$,   Eq. (\ref{opt-symm19}) is  fulfilled  if $w=1$, that is, if $Q = 0$. This example belongs to the special case where the density operators resolve the identity operator, as discussed in Sec.  V A. 

The preceding calculations have been performed assuming that  $|b_2|\geq |b_1|$.  Since $Q$ has to be positive, we conclude from Eqs. (\ref{opt-symm20}) and (\ref{opt-symm21})  that this assumption is justified if  $|c_2|\leq |c_1|$.
In the opposite case,  $|c_2|\geq |c_1|$,  the  derivation could be performed in a completely analogous way, where the results arise from the  previous ones when the indexes 1 and 2 are interchanged and when $m$ is replaced by $d-m$, and vice versa.


\begin{thebibliography}{99}

\bibitem{helstrom}
    C. W. Helstrom, {\it Quantum Detection and Estimation Theory}
    (Academic Press, New York, 1976).

\bibitem{holevo}A. S. Holevo, {\it Probabilistic and Statistical Aspects of Quantum Theory}
(North-Holland, Amsterdam, 1979).  

\bibitem{ivan} I. D. Ivanovic, Phys. Lett. A {\bf 123}, 257
  (1987).

\bibitem{dieks} D. Dieks, Phys. Lett. A {\bf 126}, 303 (1988).

\bibitem{peres} A. Peres, Phys. Lett. A {\bf 128}, 19 (1988).

\bibitem{jaeger} G.\ Jaeger and A.\ Shimony, Phys.\ Lett. A {\bf 197},
  83 (1995).

\bibitem{rudolph} T. Rudolph, R. W. Spekkens, and P. S. Turner,
  Phys. Rev. A {\bf 68}, 010301(R) (2003).

\bibitem{raynal} Ph. Raynal, N. L\"utkenhaus, and S. van Enk,
 Phys. Rev. A {\bf 68}, 022308 (2003).

\bibitem{eldar}Y. C. Eldar, M. Stojnic, and B. Hassibi, \pra {\bf 69},
062318 (2004).

\bibitem{HB} U. Herzog and J. A. Bergou, \pra {\bf 71},
050301(R) (2005).

\bibitem{BFH}
J. A Bergou, E. Feldman, and M. Hillery, \pra {\bf 73}, 032107
(2006).

\bibitem{herzog} U. Herzog, \pra {\bf 75}, 052309 (2007).

\bibitem{raynal2} Ph. Raynal and N. L\"utkenhaus, \pra {\bf 76}, 052322 (2007).

\bibitem{kleinmann} M. Kleinmann, H. Kampermann, and D. Bru\ss,
Phys. Rev. A {\bf 81}, 020304(R) (2010).

\bibitem{chefles1} A. Chefles, Phys. Lett. A {\bf 239}, 339 (1998).

\bibitem{croke} S. Croke, E. Andersson, S. M. Barnett, C. R. Gilson
and J. Jeffers, Phys. Rev. Lett. {\bf 96}, 070401 (2006).

\bibitem{mosley} P. J. Mosley, S. Croke, I. A. Walmsley and S. M.
Barnett, \prl {\bf 97}, 193601 (2006).

\bibitem{herzog1} U. Herzog, Phys. Rev. A {\bf 79}, 032323 (2009).

\bibitem{herzog-benson} U. Herzog and O. Benson,  J. Mod. Opt. {\bf 57}, 188
(2010).

\bibitem{steudle} G. A. Steudle, S. Knauer, U. Herzog, E. Stock, V.
Haisler, D. Bimberg, and O. Benson, Phys. Rev. A {\bf 83}, 050304(R) (2011).

\bibitem{jimenez}  O. Jim\'enez, M. A. Solis-Prosser, A. Delgado, and L. Neves,
Phys. Rev. A {\bf 84}, 062315  (2011).

\bibitem{herzog2} U. Herzog,  Phys. Rev. A {\bf 85}, 032312 (2012).

\bibitem{chefles-barnett}  A. Chefles and S. M. Barnett,
J. Mod. Opt. {\bf 45}, 1295 (1998).

\bibitem{zhang-li}
C. W. Zhang, C. F. Li, and G. C. Gou, Phys. Lett. A {\bf 261}, 25 (1999).

\bibitem{fiurasek}
J. Fiur\'a\v{s}ek and M. Je\v{z}ek, \pra 67, 012321 (2003).

 \bibitem{eldar1}
 Y. C. Eldar, \pra {\bf 67}, 042309 (2003).

\bibitem{touzel}
M. A. P. Touzel, R. B. A. Adamson, and A. M. Steinberg, \pra {\bf
76}, 062314 (2007).

\bibitem{hayashi-err}
A. Hayashi,  T. Hashimoto, and M. Horibe,  \pra {\bf 78}, 012333 (2008).

\bibitem{sugimoto}
H. Sugimoto, T. Hashimoto, M. Horibe, and A. Hayashi, \pra {\bf 80},
052322 (2009).

\bibitem{neumark} M.\ A.\ Neumark,    
Izv. Akad. Nauk SSSR, Ser. Mat. {\bf 4}, 277 (1940).    
   
\bibitem{preskill}  J. Preskill,    
{\it Lecture Notes for Physics 229: Quantum Information and Computation}    
(Cambridge University Press, 1998).   

\bibitem{ban}  M. Ban, K. Kurokawa, R. Momose, and O. Hirota, Int. J.
Theor. Phys. {\bf 36}, 1269 (1997).

\bibitem {chefles-symm} A. Chefles and S. M. Barnett, Phys. Lett. A {\bf 250}, 223
(1998).

\bibitem{clarke} R. B. M. Clarke, V. M. Kendon, A. Chefles, S. M. Barnett, E. Riis, and M. Sasaki, 
\pra {\bf 64}, 012303 (2001).

\bibitem{yuen}H. P. Yuen, R. S. Kennedy, and M. Lax, IEEE Trans. Inf. Theory {\bf 21}, 125 (1975). 

\bibitem{jafarizadeh} M. A. Jafarizadeh, Y. Mazhari, and M. Aali, Quantum Inf. Process {\bf 10},  155 (2011).

\bibitem{kimura} G. Kimura, T. Miyadera, and H. Imai,   \pra {\bf 79}, 062306 (2009).

\bibitem{bagan} E. Bagan, R. $\rm Mu\tilde{n}oz$-Tapia,  G. A. Olivares-Renter\'{i}a 
  and J. A. Bergou,
arXiv:1206.4145 [quant-ph].


\end{thebibliography}
\end{document}